\documentclass[twocolumn,showpacs,preprintnumbers,amsmath,amssymb]{revtex4}
\usepackage{dcolumn}
\usepackage{bm}
\usepackage{latexsym}
\usepackage{amsfonts}
\usepackage[dvips]{graphicx}
\usepackage[usenames]{color}
\usepackage{makeidx}

\newcommand{\ket}{\rangle }
\newcommand{\bra}{\langle }

\newcommand{\ve}{\varepsilon}

\newcommand{\up}{\uparrow}
\newcommand{\dw}{\downarrow}

\newcommand{\Vect}[1]{\mbox{\boldmath$#1$}}

\def\widebar{\accentset{{\cc@style\underline{\mskip10mu}}}}

\begin{document}
\title{
Nonlinear doublon production in a Mott insulator\\
--- Landau-Dykhne method applied to an integrable model
}
\author{Takashi Oka}
\affiliation{Department of Physics, University of Tokyo, Hongo, Tokyo 113-0033, 
Japan}
\affiliation{Department of Physics, Harvard University, Cambridge, MA 02138, USA}
\date{\today}
\begin{abstract}
\noindent 
Doublon-hole pair production which 
takes place during dielectric breakdown in 
a Mott insulator subject to a strong
laser or a static electric field is studied
in the one-dimensional Hubbard model. 
Two nonlinear effects cause the excitation, i.e.,
multi-photon absorption and
quantum tunneling.
Keldysh crossover between the two mechanisms occurs
as the field strength and photon energy is changed. 
The calculation is done analytically 
by the Landau-Dykhne method in combination with the Bethe ansatz solution
and the results are compared with those of the time dependent 
density matrix renormalization group. 
Using this method, we calculate 
distribution function of the generated doublon-hole pairs 
and show that it drastically changes as we cross the 
Keldysh crossover line. 
After calculating the tunneling threshold for several
representative one-dimensional Mott insulators, 
possible experimental tests of the theory is proposed
such as {\it angle resolved photoemission spectroscopy 
of the upper Hubbard band} in the quantum tunneling regime. 
We also discuss the relation of the present theory with 
a many-body extension of 
electron-positron pair production in
nonlinear quantum electrodynamics known as the Schwinger mechanism. 

\end{abstract}

\maketitle
\section{ Introduction}

``Nonequilibrium strongly correlated systems" 
is becoming an important field of study in condensed matter physics
\cite{Iwai2003,Wall2011,Tokura88,tag,Moriheating,Watanabe2011,
HiroriAPL2011,LiuNat2012,
Greiner2002,StrohmainerEsslinger10,GreifEssslinger2011,Kollath06prl}. 
These systems offer a testbed for theoretical advances
such as the extension of the linear response 
paradigm to nonlinear processes \cite{Oka2003,Oka2005a,
OkaSchwinger10,Oka2005b,PhysRevLett.105.146404,Eckstein2011,Tanaka2011,
Heidrich-Meisner10,Takahashi08B,OkaPhotoLuttinger,Oka2004a,SachdevElectricField02,Ajisaka2009,Nakumura2010}. 
We can experimentally induce a nonequilibrium state in 
photo-induced phase transitions in solids 
\cite{Iwai2003,Wall2011}
as well as in the dynamics of cold atoms 
\cite{Greiner2002,StrohmainerEsslinger10,GreifEssslinger2011,Kollath06prl}. 
The photo-induced insulator to metal transition in Mott insulators has 
generated  substantial interest because it is one of the most basic nonequilibrium phenomena in strongly correlated systems \cite{Iwai2003,Wall2011} . 
The response of Mott insulators subject to strong external fields 
has been studied experimentally. 
Initially, the motion of particles (electrons or atoms) is frozen by strong repulsion, and the ground state is a 
Mott insulator \cite{Imada1998}. 
Perturbations (electric field or 
lattice modulation) excite pairs 
of doublons (= doubly occupied site) 
and holes (= sites with no electrons; 
we do not call this state ``holon" because this is a  Bethe ansatz terminology 
that is used later), and
when their density becomes sufficiently
high, ``melting" of the Mott state occurs 
leading to an insulator-to-metal transition\cite{Iwai2003}. 
Quite recently, the insulator-to-metal transition was
realized by a terahertz laser in vanadium dioxide, which is a 
candidate material for a Mott transition \cite{LiuNat2012}. 
Since the photon energy is far below the optical gap, 
the excitation mechanism is expected to 
be a nonlinear process. 
These experiments give us strong motivation to develope a 
theory for nonlinear excitations in strongly correlated
systems.

The purpose of this study is to gain an analytical understanding 
of the excitation process when a
strong electric field is applied to a Mott insulator. 
The effects of strong electric fields 
on Mott insulators have been studied extensively in
theory via the 
fermionic Hubbard model 
using numerical methods such as exact diagonalization
\cite{Oka2003,Takahashi08B}, the time-dependent density matrix renormalization 
group (td-DMRG) \cite{Oka2005a,OkaSchwinger10,Heidrich-Meisner10},
and nonequilibrium dynamical mean field theory 
\cite{PhysRevLett.105.146404,Eckstein2011}. 
These studies reveal the following consensus. 
Doublon-hole pairs (dh-pairs) are created by strong electric fields,
and for DC-electric fields, production rate (or 
ground state decay rate)
shows a threshold behavior \cite{Oka2003,Oka2005a}. 
This behavior seems to be universal and independent 
of dimensions, e.g.,
Refs.~\cite{Oka2003,Oka2005a,OkaSchwinger10,Heidrich-Meisner10}
(one-dimensional (1-D) studies) and Ref.~\cite{PhysRevLett.105.146404} 
(infinite-dimensional studies).
If we denote tunneling threshold by $F_{\rm th}$,
for small electron repulsion $U$, 
it behaves as $F_{\rm th}\propto \Delta_{\rm Mott}^2$,
where $\Delta_{\rm Mott}$ is the Mott gap.
We can obtain this expression by applying the Landau-Zener formula 
to many-body energy levels\cite{Oka2003}. 
For AC-electric fields, 
it was mentioned in Ref.~\cite{Takahashi08B}
that there is a crossover from a weakly excited state
to a strongly excited state with increasing 
field strength. 
Another interesting observation was made regarding the
bosonic Hubbard model with lattice modulation, 
where the authors calculated energy absorption rate using
td-DMRG \cite{Kollath06prl}. 
The absorption peaked around $\Omega\sim N U$ 
($\Omega$: modulation frequency, $N$: integer) 
broadened as modulation intensity increased. 
The broadening is clearly a nonlinear effect.

In this study, we examine the 1D Hubbard model
at half-filling and the instability of the 
ground state in strong electric fields. 
The method we use is a combination of the
Landau-Dykhne quantum tunneling theory 
\cite{Dykhne1962,LandauLifshitzQM,DavisPechukas1976} and the Bethe ansatz.
This method was developed in Ref.~\cite{OkaSchwinger10}
and was used to derive an analytic expression for tunneling threshold. 
Although not commonly employed in condensed matter, 
the Landau-Dykhne method 
has a long application history in areas of physics 
where quantum systems are driven out
of their initial state by strong external fields. 
The name ``strong field physics" is often used to describe this 
problem area in physics.  
We can find examples of driven systems in quantum chemistry
\cite{DavisPechukas1976},
atom ionization \cite{Keldysh65,DeloneKrainovBook},
quantum chaos \cite{Wilkinson2000},
and high energy 
\cite{Heisenberg1936,Schwinger1951,PopovSovJNP1974,PopovJETP1972,Brezin1970,Popovreview,DumluDunne10,Dunne04review}.
``Nonlinear excitations in Mott insulators"
is a typical problem in ``strong field physics in condensed matter"
(for a review, see \cite{Okareview,Okareview2}).
Ideas and techniques developed in other fields prove 
quite useful as well. 

Our problem has many common features with the 
electron-positron pair production problem 
in nonlinear quantum electrodynamics (QED)
(for a review, see Ref.~\cite{Dunne04review}). 
The concept of vacuum in high energy physics 
is directly translated into an ``insulator" in condensed matter. 
Shortly after the report of the Dirac sea vacuum description, 
Heisenberg-Euler 
proposed that nonlinear response of the vacuum 
is described by an effective Lagrangian \cite{Heisenberg1936}.
They also found vacuum instability against 
electron-positron pair production
when field strength is comparable to tunneling threshold.
This threshold is now called the Schwinger limit \cite{Schwinger1951}. 
The calculation of production rate 
was extended from DC to AC electric
fields\cite{Brezin1970,PopovSovJNP1974,PopovJETP1972}
following an early study by Keldysh on
atom ionization \cite{Keldysh65}. 
In particular,
Popov used the Landau-Dykhne method
to calculate production rate 
\cite{PopovSovJNP1974,PopovJETP1972} 
(for a review of this approach, see Ref.~\cite{Popovreview}). 
Following these ``strong field physics"-studies, 
a universal picture emerged that, in fact, had already been 
noticed by Keldysh\cite{Keldysh65}.
That is, there are two leading excitation mechanisms 
in a zero-temperature-gapped system 
driven by an AC-field. 
One is quantum tunneling, which is dominant in the 
DC limit, and shows a threshold behavior.
This threshold is nothing but the Schwinger limit.
The other mechanism is multi-photon absorption that 
is dominant when photon energy is relatively large.
Moreover, production rate shows a power law behavior. 
There is a crossover between the two regimes, 
which is called the Keldysh crossover.

\begin{figure}[tb]
\centering 
\includegraphics[width=8cm]{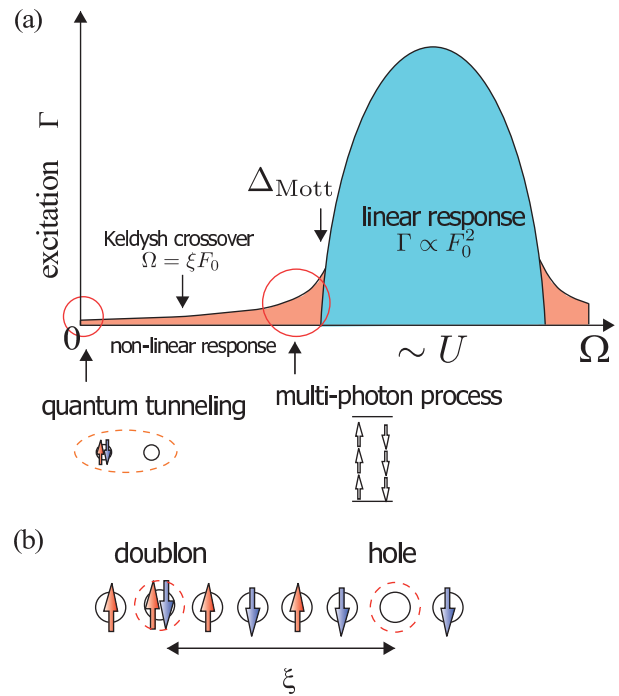}
\caption{(color online)
(a) Schematic plot of the nonlinear optical absorption spectrum 
of a Mott insulator in a strong AC-electric field.
$F_0(>\; 0)$ is the field strength and $\Omega$ is the photon energy of the
applied laser.
In addition to the contribution from 
linear response theory (Kubo formula),
sub-gap excitations occur owing to nonlinear processes. 
Mechanisms are quantum tunneling and multi-photon absorption. 
They are governed by the Mott gap $\Delta_{\rm Mott}$ and 
correlation length $\xi$.
(b)
Correlation length $\xi$ 
is the typical size of doublon-hole pairs 
in the Mott insulating ground state. 
Pair production and annihilation occur
during the virtual process. 
}
\label{fig:schematic}
\end{figure}

In this study, we show that the nonlinear dh-excitation
in a Mott insulator also lies within the Keldysh paradigm. 
Using the Landau-Dykhne method combined with the Bethe ansatz, we
derive an expression for the momentum-resolved production
rate of dh-pairs (Eq.~(\ref{eq:LD2}) below)
and calculate the total production rate $\Gamma$. 
In Fig.~\ref{fig:schematic}, we schematically plot 
the production rate behavior in strong AC-electric fields
\begin{eqnarray}
F(t)=F_0\sin\Omega t.
\end{eqnarray}
Here, $F_0=eaE(>0)$ is the field strength,
$e$ is the electron charge, $a$ is
the lattice constant, and $\Omega$
is the photon energy.
When photon energy is above the Mott gap and is resonant 
with the absorption spectrum, i.e., $\Omega\sim U$, 
we obtain the standard linear response result, i.e.,
$\Gamma\propto F_0^2$.
In the case where the field is off-resonant $\Omega<\Delta_{\rm Mott}$, 
nonlinear processes lead to dh-production. 
Similar to the other above-mentioned ``strong field physics"
 examples,
the two leading mechanisms 
are multi-photon absorption and quantum tunneling. 
The former occurs when $\Omega$ is relatively 
close to the gap. Production rate has the following power law 
dependence on field strength
\begin{eqnarray}
\Gamma\propto \left(
\frac{F_0\xi}{2\pi\Omega}
\right)^{2\frac{\Delta_{\rm Mott}}{\Omega}}.
\end{eqnarray}
Here, the power $2\frac{\Delta_{\rm Mott}}{\Omega}$
is twice the number of absorbed photons
and the factor $\xi$ 
is the doublon-hole correlation length \cite{Stafford1993}. 
In the ground state of a Mott insulator, 
doublon-hole pairs are created during
a quantum mechanical virtual process 
[Fig.~\ref{fig:schematic}(b)].  
Correlation length gives the typical size of such doublon hole pairs. 
When the DC limit is approached with a small $\Omega$, 
the leading mechanism becomes quantum tunneling.
This leads to a dielectric breakdown with a 
threshold behavior \cite{Oka2003,Oka2005a,OkaSchwinger10,PhysRevLett.105.146404}. 
The total production rate in this regime has the 
approximate form
\begin{eqnarray}
\Gamma\propto
\exp\left(-\pi\frac{F_{\rm th}}{ F_0}\right),
\end{eqnarray}
where DC threshold (=Schwinger limit) is given by
\begin{eqnarray}
F_{\rm th}\sim \frac{\Delta_{\rm Mott}}{2\xi}.
\label{eq:Schwingerlimit0}
\end{eqnarray}
We notice that the correlation length $\xi$ again
plays an important role. 

The paper is organized as follows.
In section \ref{section:method}, after a brief introduction
of the Bethe ansatz solution of the Hubbard model, we explain how to combine
its information with the Landau-Dykhne method. 
Application to nonlinear transport in DC fields
and photo-induced phase transitions in AC fields
is discussed in sections \ref{sec:DC} and 
\ref{sec:AC}, respectively. 
In section \ref{sec:exp}, we discuss experimental feasibility.

\begin{figure*}[hbt]
\centering 
\includegraphics[width=13cm]{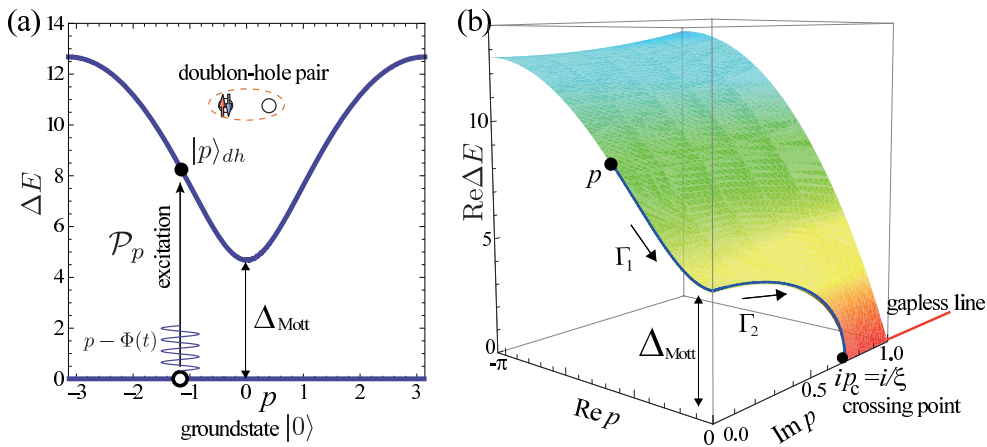}
\caption{(color online)
(a) The excitation energy $\Delta E$ 
of the doublon-hole pair as a function of 
the hole momentum $p$
in the $U=8$ Hubbard model. 
$\mathcal{P}_p$ is the tunneling probability 
to create a state $|p\ket_{dh}$ from $|0\ket$. 
(b)
The real part of the excitation energy 
plotted for complex $p$. 
The $\mbox{Im}p=0$ slice is equivalent to the left half of (a). 
The gap closes at the level crossing point $p = ip_{\rm c}$ 
where a gapless line starts. 
Paths $\Gamma_1$ and $\Gamma_2$ are  
used in the integral in Eq.~(\ref{eq:ImDHubbard}).
}
\label{fig1}
\end{figure*}

\section{Landau-Dykhne + Bethe ansatz method}
\label{section:method}
In this section, we extend the Landau-Dykhne + Bethe ansatz method, 
developed in Ref.~\cite{OkaSchwinger10}, to 
electric fields with various laser types. 
The model we study is the half-filled 1D fermionic Hubbard model subject to an electric field. 
The Hamiltonian is given by 
\begin{eqnarray}
H(\Phi)&=&-\tau\sum_{j,\sigma}(e^{i\Phi}c^\dagger_{j+1\sigma}c_{j\sigma}+e^{-i\Phi}c^\dagger_{j\sigma}c_{j+1\sigma})
\label{eqn:HubbardHamiltonian}\\
&&+U\sum_jn_{j\up}n_{j\dw}\nonumber.
\end{eqnarray}
The time-dependent Peierls phase $\Phi$ is related to the applied
electric field by $F(t) = eaE(t) = -d\Phi(t)/dt$. 
We set the energy unit as the hopping 
amplitude, i.e., $\tau=1$.  
We start from the Mott insulating ground state at $t=0$
and apply the electric field for $t > 0$.

The static Hubbard model can be solved exactly
using the Bethe ansatz
and the ground state wave function 
as well as
excitations is well understood\cite{Hubbardbook}. 
There are two types of elementary excitations
from the half-filled ground state: 
(1) Gapped spinless excitations
with charge $\mp e$ called antiholons and holons
(2) Gapless charge neutral excitations carrying spin 
$\pm \frac{1}{2}$ called spinons. 
Physical excitations are built from these
elementary excitations. 
In the remainder of this article, 
instead of using the Bethe ansatz terminology 
antiholon and holon, we use the more familiar names
doublon and hole. 

Among the excitations, we concentrate on excited states with a single doublon-hole 
pair, i.e., antiholon-holon pairs. 
The states are parameterized
by rapidity $k_1$ ($k_2$) for a hole (doublon).
The total energy and central momentum of these excitations are given by
\begin{eqnarray}
\Delta E\equiv E_{dh}-E_0&=&\ve_h(k_1)+\ve_{d}(k_2),
\label{eq:EBethe}\\
P_{\rm central}&=&p_h(k_1)+p_{d}(k_2),
\end{eqnarray}
where $\ve_{h,d}$ and $p_{h,d}$ are
the energy and momentum of the 
hole and doublon, respectively.
Holon energy is given by\cite{Hubbardbook} 
\begin{eqnarray}
\ve_h(k)&=&\ve_{d}(k)=U/2+2\cos k\nonumber\\
&&+2\int_0^\infty\frac{d\omega}{\omega}
\frac{J_1(\omega)\cos(\omega \sin k)e^{-U\omega/4}}
{\cosh(\omega U/4)}
\end{eqnarray}
and holon momentum is given by 
\begin{eqnarray}
p_h(k)&=&p_{d}(k)+\pi
=\frac{\pi}{2}-k
\nonumber\\
&&-2\int_0^\infty\frac{d\omega}{\omega}
\frac{J_0(\omega)\sin(\omega\sin(k))}{1+\exp (U\omega/2)}.
\end{eqnarray}
Note that we shifted holon momentum by $\frac{\pi}{2}$, 
i.e., $p\equiv p_h+\frac{\pi}{2}(=
-p_{d}-\frac{\pi}{2})$.
Only states with $P_{\rm central}=0$ can be excited by an 
external electric field because
the momentum of the laser can be ignored.
We denote a single doublon-hole pair with the
hole momentum $p$ by $|p\ket_{dh}$.
In Fig.~\ref{fig1}(a), we plot the excitation energy
$\Delta E(p)$ as a function of $p$. 
It has a minimum at $p=0$
with a gap $\Delta_{\rm Mott}$, i.e., the Mott gap. 

\begin{figure}[htb]
\centering 
\includegraphics[width=7cm]{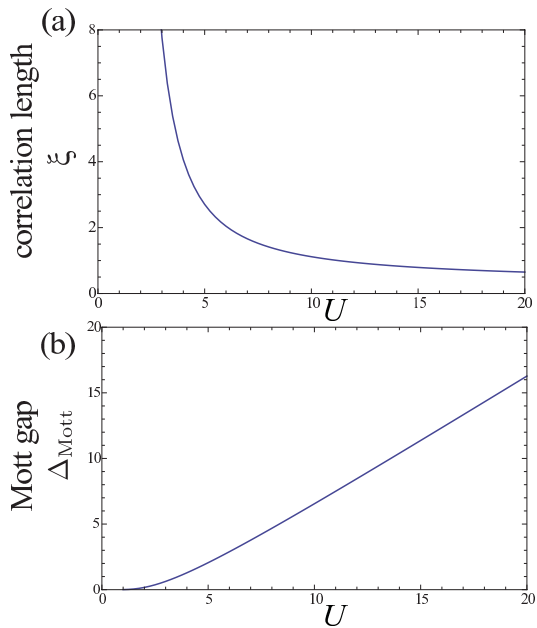}
\caption{(color online)
(a) Correlation length $\xi$ \cite{Stafford1993}
and
(b) Mott gap $\Delta_{\rm Mott}$ (Lieb-Wu solution)
of the 1D Hubbard model at half-filling. 
}
\label{fig:xiDelta}
\end{figure}

An important concept in the Mott insulating phase
is the correlation length $\xi$ studied by 
Stafford and Millis in Ref.~\cite{Stafford1993}
using the Bethe ansatz.
In Mott insulators, each lattice site is 
occupied by a single electron. 
However, quantum fluctuations 
enable doublons and holes to pair create and wander around as a virtual process before they pair annihilate.
This process is responsible for 
antiferromagnetic super exchange coupling.
Correlation length is intuitively the size of the 
doublon-hole pair in the ground state wave 
function [Fig.~\ref{fig:schematic}(b)]. 
If we consider a finite system of size $L$,
charge stiffness (Drude weight) decays as
\begin{eqnarray}
D_c(L)\propto \exp(-L/\xi).
\end{eqnarray}
In other words, if the system is small enough
compared to $\xi$, it behaves as a metal
because the carriers (doublons and holes)
can transport current. 
Another important property is 
that Green's function 
$G(|x-x'|)=\bra 0|c^\dagger_{x'\sigma}c_{x\sigma}+
c^\dagger_{x\sigma}c_{x'\sigma}|0\ket$
decays as follows \cite{Stafford1993}:
\begin{eqnarray}
G(|x|)\sim\exp(-|x|/\xi).
\label{eq:greendecay}
\end{eqnarray}
The exact expression for $\xi$ is given by\cite{Stafford1993}
\begin{eqnarray}
1/\xi=\frac{4}{U}\int_1^\infty dy\frac{\ln(y+\sqrt{y^2-1})}
{\cosh(2\pi y/U)}.
\end{eqnarray}
In Fig.~\ref{fig:xiDelta}(a), we plot $\xi$
as well as $\Delta_{\rm Mott}$
as a function of $U$.
In the small $U$ limit, it behaves as
\begin{eqnarray}
\lim_{U\to 0}\xi=\frac{2t+U/2\pi+\cdots}{\Delta(U,t)} 
\end{eqnarray}
whereas the strong-coupling limit is given by
\begin{eqnarray}
\xi^{-1}=\ln(U/at)\quad U\to \infty
\end{eqnarray}
with $a=[\Gamma(1/4)/\sqrt{2\pi}]^4\simeq 4.377 $.

Next, we consider time evolution. 
We start from the ground state $|0\ket$
and apply an electric field described by a 
time-dependent phase via $F(t) = eaE(t) = -d\Phi(t)/dt$. 
After a tunneling process, 
the wave function takes the form 
\begin{eqnarray}
|\Psi\ket\sim e^{-i\alpha}\sqrt{\mathcal{N}}|0\ket+\sum_p\sqrt{\mathcal{P}_p}e^{-i\beta_p}|p\ket_{dh}+\ldots,
\end{eqnarray}
 ($\alpha,\;\beta_p$ are phases)
where the ground state amplitude decreases
as $\sqrt{\mathcal{N}}\sim\prod_p\sqrt{1-\mathcal{P}_p}$\cite{Oka2005a}.
$\mathcal{P}_p$ is the momentum resolved 
tunneling probability of the doublon-hole pair with 
momentum $p$ [Fig.~\ref{fig1} (a)]. 
The omitted term ``$\ldots$" contains
excitations to states with multiple doublon-hole pairs as well 
as spin excitations.

One can calculate the tunneling probability $\mathcal{P}_p$ 
in the Hubbard model by 
the Landau-Dykhne method \cite{OkaSchwinger10}. 
The Landau-Dykhne method \cite{Dykhne1962,DavisPechukas1976}
(for a textbook and useful reference see 
Ref.~\cite{DeloneKrainovBook} and \cite{Vasilev2004},
respectively) has been derived from the 
adiabatic perturbation theory.
We denote the adiabatic eigenstates 
of $H(\Phi)$ by
$|0;\Phi\ket$ and $|p;\Phi\ket_{dh}$
where
\begin{eqnarray}
H(\Phi)|0;\Phi\ket&=&E_0(\Phi)|0;\Phi\ket,\\
\;H(\Phi)|p;\Phi\ket_{dh}&=&E_{dh}(p;\Phi)|p;\Phi\ket_{dh}
\end{eqnarray}
is satisfied. 
Because $p$ is a good quantum number, 
states with different $p$ are orthogonal to each other. 
Thus, by ignoring multiple pair states, 
we can study each excitation channel independently.
This means that we can study the tunneling process in a Hilbert space 
spanned by two states,
i.e., 
the problem reduces to solving the time-dependent Schr\"odinger
equation with a solution of the form
\begin{eqnarray}
|\Psi(t)\ket=a(t)|0;\Phi(t)\ket+b(t)|p;\Phi(t)\ket_{dh}
\end{eqnarray}
(initial condition $a(0)=1,\;b(0)=0$). 
This significantly simplifies the problem.
Landau-Dykhne's tunneling theory 
states that the 
tunneling probability ($|b(t_{\rm end})|^2$)
between two quantum levels is given by
\begin{eqnarray}
\mathcal{P}_p=\exp(-2\mbox{Im}\mathcal{D}_p),
\label{eq:LD}
\end{eqnarray}
where 
\begin{eqnarray}
\mathcal{D}_p=\int_\gamma \left[E_{dh}(p;\Phi(t))-E_0(\Phi(t))\right]dt
\end{eqnarray}
is the difference between the dynamical phase
of the ground state and the excited state. 
In our problem, this expression can be simplified 
because the effect of the Peierls phase on the 
adiabatic solutions is expressed simply by replacing 
the momentum $p$ by $p-\Phi$.
Thus, we have 
\begin{eqnarray}
|p;\Phi\ket_{dh}=|p-\Phi\ket_{dh},\;
 E_{dh}(p;\Phi)=E_{dh}(p-\Phi),
\end{eqnarray} 
which leads to 
$
\mathcal{D}_p=\int_\gamma \Delta E(p-\Phi(t))dt,
$
where $\Delta E$ is defined in Eq.~(\ref{eq:EBethe}).

An interesting point of the Landau-Dykhne formula [Eq.~(\ref{eq:LD})]
is that tunneling probability depends on the 
{\it imaginary} part of the dynamical phase difference. 
Integration path $\gamma$ starts from $t=0$ and ends at a
critical time $t=t_{c}$, where level crossing takes place. 
Because we are dealing with a gapped system, 
level crossing does not occur for real $t$; instead, it 
occurs at a complex time when
\begin{eqnarray}
\Delta E(p-\Phi(t_c))=0
\label{eq:levelcrossing}
\end{eqnarray}
is satisfied. 
When $t$ is a complex number, the corresponding
Peierls phase is also complex. 
The Hubbard Hamiltonian [Eq.~(\ref{eqn:HubbardHamiltonian})]
with a complex Peierls phase 
is a non-Hermitian lattice model where the 
absolute values of the left and right hoppings are unequal. 
The ground state wave function of the non-Hermitian model
was studied by Fukui and Kawakami in ref.~\cite{Fukui}.
In Fig.~\ref{fig1}(b), we plot 
$\mbox{Re}\Delta E$ for complex $p$.
The level crossing is found at a point $ip_c\equiv p-\Phi(t_c)$. 
The momentum of the 
level crossing point is related to correlation length by
\cite{Stafford1993,NakamuraHatano06}
\begin{eqnarray}
p_c=1/\xi,
\end{eqnarray}
i.e., they are the inverse of each other. 
In a noninteracting system, this is a very natural relationship,
which states that localization length is the inverse of 
complex momentum and a wave function decays as $e^{ikx}=e^{-\kappa x}$
when $k=i\kappa$. However, the surprise here is that this concept can be 
extended to a many-body system in a straightforward manner.

The expression for tunneling probability becomes 
physically clearer when we change variables 
in the integral from time $t$ to the Peierls phase $\Phi$. 
Using the Jacobian $\frac{dt}{d\Phi}=-1/F$, where $F$ 
is the electric field expressed as a function of $\Phi$,
we are led to the expression
\begin{eqnarray}
\mathcal{P}_p=\exp\left(-2\mbox{Im}\int_0^{\Phi_{\rm c}}
\Delta E(p-\Phi)\frac{-1}{F(\Phi)}
d\Phi\right),
\label{eq:LD2}
\end{eqnarray}
which is the main result of this study. 
We keep the minus sign in Eq.~(\ref{eq:LD2})
as a reminder that the factor in the exponential 
is negative. 
This expression is a direct descendant of 
V.~S.~Popov's original expression for the tunneling probability 
in the massive Dirac model\cite{PopovJETP1972}. 
The difference here is that the one-body energy level is replaced
by the many-body level obtained by the Bethe ansatz. 
One can use Eq.~(\ref{eq:LD2}) to study 
excitations not only for DC-fields but also
for various other fields. This can be achieved by simply replacing the
function $F(\Phi)$ in the formula.
Table \ref{table:Jacobian} shows the models of the 
electric fields
we use in this study. 

\begin{table}[tb]
\centering
  \caption{Models of electric fields. 
$F_0$ is the field strength, $\Omega$ the 
photon energy and $\sigma$ the pulse duration. 
}
   \begin{tabular}{cccc}
\hline
type&$F(t)$&$F(\Phi)$&attempt frequency $f$\\
\hline
DC-field&$F_0$&$F_0$&$F_0/2\pi$\\
AC-field&$F_0\sin\Omega t$&$
\pm\sqrt{F_0^2-\Omega^2\Phi^2}$&$\Omega/2\pi$\\
single pulse&$F_0\cosh^{-2}(t/\sigma)$&$F_0\left(1-\frac{\Phi^2}{\sigma^2F_0^t}\right)$&1(single process)\\
\hline
   \end{tabular}
   \label{table:Jacobian}
\end{table}
There is an arbitrariness in the
$\Phi$ integral in Eq.~(\ref{eq:LD2}). 
Because $\mathcal{D}_p$
is a complex integral of an analytic function, one can 
deform its path as long as the end points
are fixed and no singular point is crossed. 
The most natural path that simplifies
the calculation is the one 
that goes from $p$ to $0$ ($\Gamma_1$)
and then from $0$ to $ip_{\rm c}$ ($\Gamma_2$) 
as shown in Fig.~\ref{fig1}(b).
On this path, $\Delta E$ is always real. 
The integral is divided into two 
\begin{eqnarray}
\mathcal{D}_p=\int_0^{\Phi_{\rm c}}\Delta E(p-\Phi)\frac{-1}{F(\Phi)}
d\Phi=\mathcal{D}_{p1}+\mathcal{D}_{p2},
\label{eq:ImDHubbard}
\end{eqnarray}
corresponding to $\Gamma_1$ and $\Gamma_2$,
and the two contributions 
can be written as real integrals
\begin{eqnarray}
\mbox{Im}\mathcal{D}_{p1}=\int_p^0\Delta E(l)\mbox{Im}
\left(\frac{1}{F(p-l)}\right)dl,
\label{eq:Dp1}
\\
\mbox{Im}\mathcal{D}_{p2}=\int_0^{p_{\rm c}}\Delta E(il)
\mbox{Im}\left(\frac{1}{F(p-il)}\right)dl.
\label{eq:Dp2}
\end{eqnarray}
In these integrals, the variable $l$ is real. 
We can numerically perform the integration, and in 
some simple cases, we can derive analytical expressions, as 
we will see later. 

We now comment on the relationship between 
tunneling probability and production rate. 
The tunneling probability $\mathcal{P}_p$ 
is defined for a single excitation attempt. 
To promote it to a ``rate", we must multiply it with the 
attempt frequency $f$, i.e., the number of events per unit time.
Then, momentum resolved production rate is given by
\begin{eqnarray}
\Gamma_p=f\mathcal{P}_p.
\end{eqnarray}
In the case of DC fields, 
physical momentum, defined on the interval $(-\pi,\pi]$, 
evolves as $p-\Phi(t)=p+F_0t$.
Thus, the time period between tunneling events 
is $2\pi/F_0$:, which is nothing but the 
period of the Bloch oscillation. 
In this case, the attempt frequency 
is given by its inverse $f=F_0/2\pi$. 
As for the AC field with photon energy $\Omega$,
the attempt frequency is $f=\Omega/2\pi$.
When we consider a single pulse, 
we simply set $f=1$.
Table \ref{table:Jacobian} 
summarizes attempt frequency and the 
Jacobian for several models of electric fields. 

The total production rate is defined by
\begin{eqnarray}
\Gamma=f\int_{-\pi}^\pi\frac{dp}{2\pi} \mathcal{P}_p.
\label{eq:totalproductionrate}
\end{eqnarray}
The total production rate is an important quantity
because it is comparable to quantities obtained from
with other methods such as td-DMRG.
First, the total production rate gives the
lowest order approximation for the  
ground state decay rate \cite{Oka2005a}. 
The ground-state-to-ground-state 
transition amplitude (fidelity amplitude) 
for a time-dependent Hamiltonian is defined as
\begin{eqnarray}
\Xi (t)
&=&\bra 0;\Phi(\tau)|\hat{T}e^{-\frac{i}{\hbar}
\int_0^t H(\Phi(s))ds}|0;\Phi(0)\ket
\nonumber
\\ 
&&\times e^{\frac{i}{\hbar}\int_0^\tau E_0(\Phi(s))ds},
\label{ggamplitude}
\end{eqnarray}
where $\hat{T}$ stands for time ordering. 
When the ground state is unstable in the 
external driving force, 
the absolute value of the amplitude 
decays exponentially as ($D=1$ 1D)
\begin{eqnarray}
|\Xi (t)|\sim e^{-t L^D\Gamma_{\rm g.s. decay}}.
\label{eq:decay}
\end{eqnarray}
The total production rate defined in Eq.~(\ref{eq:decay})
agrees with the ground state decay rate
up to the higher order tunneling process, i.e.,$
\Gamma_{\rm g.s. decay}=\Gamma+\ldots$, where
terms such as $(\mathcal{P}_p)^2$ are neglected. 
For example, the ground state decay rate of a band insulator 
in DC fields is given by
$\Gamma_{\rm g.s. decay}
=-F\int_{\rm B.Z.}\frac{d\Vect{k}}{(2\pi)^D}
\frac{1}{2\pi}\ln\left(1-\mathcal{P}_{\Vect{k}}\right)$
where $\Vect{k}$ is the momentum in the Brillouin zone\cite{Oka2005a}.
Expanding this with $\mathcal{P}$ to the
lowest order gives the 
total production rate of electron-hole pairs, c.f. Eq.~(\ref{eq:totalproductionrate}). 

The time evolution of the doublon density 
\begin{eqnarray}
d(t)=\frac{1}{L}\sum_{i=1}^L\bra n_{i\up}n_{i\dw}\ket(t).
\label{eq:doublondensitydef}
\end{eqnarray}
can be related to the total production rate.
As we apply a strong electric field to the 
ground state, doublon density increases
from its ground state value. 
For a continuously applied electric field, 
we assume that doublon density increases 
linearly in time, and its increase
within a time interval $\Delta t$ is given by
\begin{eqnarray}
\Delta d\sim \Delta t \Gamma. 
\label{eq:DeltadGamma}
\end{eqnarray}
Again, this expression is approximate because (a)
we ignore the production of multiple pairs
and (b) we assume that the state $|p;\Phi\ket$
has precisely one additional doublon compared with the ground state, 
which is a natural assumption when $U$ is large.

Next, we comment on the validity of our method. 
It is important to note that although we use 
an exact result (the Bethe ansatz), the calculated production rate 
is only approximate. 
One origin of error lies in the Landau-Dykhne formula itself. 
It is only valid when excitations are rare events, i.e.,
$\mathcal{P}_p\ll 1$.  
This means that we can only use Eq.~(\ref{eq:LD2})
when field strength is not too large compared
with the Schwinger limit $ F_{\rm th}$
and photon frequency is below the resonance frequency
$\Omega<\Delta_{\rm Mott}$. 
A related issue is that the 
$F_0\sim U$ resonance \cite{SachdevElectricField02}
is ignored in static electric fields. 

Another important omission is the effect of 
quantum interference between multiple tunneling events. 
This is known as the Stokes phenomenon 
and has been studied in various time-dependent problems
(e.g., Ref.~\cite{DumluDunne10,Vasilev2004,Oka2005b}).
In the present problem of a driven Mott insulator,
this is related to the pair {\it annihilation} process of 
doublon-hole pairs. 
In Ref.~\cite{Oka2005b}, the effect of pair annihilation 
and the resulting quantum interference 
was studied via mapping into an effective quantum walk. 
Quantum interference may lead to several anomalous 
behaviors. An example is dynamical localization in 
energy space \cite{Oka2005b}.
The Landau-Dykhne method ignores the interference effect,
and we will see the outcome of this later 
in Section \ref{sec:singlepulse} while presenting 
a comparison with  numerical results.

\section{Dielectric breakdown in DC fields (Schwinger limit)}
\label{sec:DC}
\begin{figure}[tbh]
\centering 
\includegraphics[width=6.5cm]{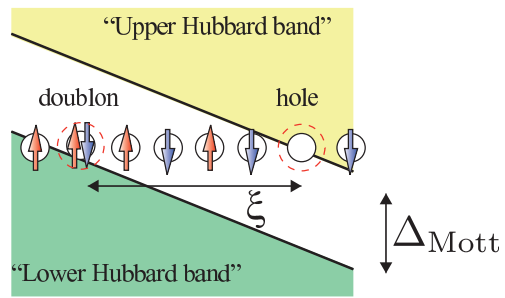}
\caption{
(color online)
Schematic rigid band description of the dielectric breakdown 
of Mott insulators in DC-electric fields. 
The upper and lower ``Hubbard bands" are tilted with 
static potential $V(x)=F_0x$ and quantum tunneling starts 
to occur when the energy drop $F_0\xi$ between the doublon pairs 
separated by $\xi$ becomes comparable to the 
excitation gap $\Delta_{\rm Mott}$. 
}
\label{fig:dcschematic}
\end{figure}

The case of DC electric fields was studied 
in Ref.~\cite{OkaSchwinger10}, and the Schwinger limit (=tunneling threshold)
was obtained and compared with td-DMRG. 
In DC fields $F = F_0$, 
when we perform the integrals in Eqs~(\ref{eq:Dp1})
and (\ref{eq:Dp2}),
we notice that the contribution
from path $\Gamma_1$ vanishes
because $F$ is always real.
Thus, we have $\mbox{Im}\mathcal{D}_p=\int_0^{p_{\rm c}}\Delta E(il)dl/F_0$
which leads to the threshold form
\begin{eqnarray}
\mathcal{P}_p=\exp\left(-\pi\frac{F_{\rm th}}{F_0}\right)
\label{eq:Schwinger0}
\end{eqnarray}
with
\begin{eqnarray}
F_{\rm th}=\frac{2}{\pi}\int_0^{p_{\rm c}=1/\xi}\Delta E(il)dl.
\label{eq:Schwinger}
\end{eqnarray} 
This coincides with earlier results \cite{OkaSchwinger10}. 
Further physical insight can be gained  
using an approximation (accurate for small $U$ \cite{footnoteE})
such as
\begin{eqnarray}
\Delta E(il)\simeq\Delta_{\rm Mott}\sqrt{1-(\xi l)^2},
\label{eq:ImE}
\end{eqnarray}
which makes the integral in Eq.~(\ref{eq:Schwinger}) trivial.
Then, we obtain the Schwinger limit 
\begin{eqnarray}
F_{\rm th}\simeq\frac{\Delta_{\rm Mott}}{2\xi}.
\label{eqn:Schwingerlimit}
\end{eqnarray}
As explained above (section \ref{section:method}), attempt frequency is given by $1/T=F_0/2\pi$,
and production rate becomes
\begin{eqnarray}
\Gamma_p= \frac{F_0}{2\pi}\exp\left(-
\pi\frac{F_{\rm th}}{F_0}\right)\qquad (\mbox{DC-fields}).
\label{eq:DC}
\end{eqnarray}
We note that the case of a DC field is special in the sense that 
production rate has no $p$-dependence. This is
because all states experience the same tunneling event.
A state with momentum $p$
drifts in the momentum space as $p+F_0 t$
and undergoes tunneling around $p+F_0 t\sim 0$
when the gap becomes smallest. 

The interpretation of the result can be simplified if we 
employ the rigid band picture (Fig.~\ref{fig:dcschematic}). 
The rigid band picture 
simply views a Mott insulator as a band insulator
with the role of conduction and valence bands 
played by the upper and lower ``Hubbard bands"
with the ``band gap" $\Delta_{\rm Mott}$. 
To make a pair with energy $\Delta_{\rm Mott}$,
the doublon and hole must be separated from each
other in a virtual process until they become
real (on-shell). The separation is on the order of
$\Delta_{\rm Mott}/F_0$ and the probability 
for this to happen is given by Green's function,
i.e.,
$\mathcal{P}\propto
G(\kappa\Delta_{\rm Mott}/F_0)$ with
$\kappa=\pi/2$. 
Because Green's function decays exponentially 
in the Mott insulating phase [Eq.~(\ref{eq:greendecay})],
we obtain a production rate
 exponentially dependent on the 
electric field, i.e., Eq.~(\ref{eq:DC}). 

If we compare Eq.~(\ref{eqn:Schwingerlimit})
with Schwinger's threshold in QED \cite{Heisenberg1936,Schwinger1951}, 
we notice that the correlation length $\xi$ 
plays the role of the 
Compton wavelength $\lambda=h/m_ec$.  
In the small $U$ limit, correlation length (soliton length) is
$\xi=2v_{\rm eff}/\Delta_{\rm Mott}$ with $v_{\rm eff}=2+U/2\pi+\cdots$ 
the speed of the charge excitations \cite{Stafford1993}. 
In this limit, we recover the Landau-Zener result \cite{Oka2003} 
\begin{eqnarray}
\Gamma_p= \frac{F_0}{2\pi}\exp\left(
-\pi\frac{(\Delta_{\rm Mott}/2)^2}{v_{\rm eff} F_0}
\right)
\quad (\mbox{DC, Small $U$})
\label{eq:LandauZener}
\end{eqnarray} 
analogous to the Schwinger mechanism in QED
with $F_{\rm th}\propto (\mbox{gap})^2$ 
\cite{Heisenberg1936,Schwinger1951}.

\begin{figure}[htb]
\centering 
\includegraphics[width=7cm]{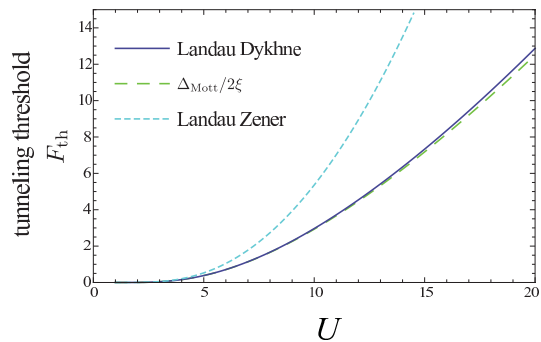}
\caption{(color online)
Schwinger limit (= tunneling threshold) \cite{OkaSchwinger10} of the 
1D Hubbard model at half-filling. 
In (c), the solid line is the Landau-Dykhne result given by 
Eq.~(\ref{eq:Schwinger}), whereas the dashed line is its approximate form
Eq.~(\ref{eqn:Schwingerlimit}). 
The Landau-Zener result in Eq.~(\ref{eq:LandauZener}) with $v_{\rm eff}=2$
is plotted as a dotted line. The Landau-Zener result is 
only accurate for a small $U$. 
}
\label{fig:threshold}
\end{figure}

In the large $U$-limit, we have
$\xi^{-1}\sim \ln(U/g\tau)\; (g \sim 4.3)$\cite{Stafford1993}
and production rate shows an interesting power law 
behavior 
\begin{eqnarray}
\Gamma_p= \frac{F_0}{2\pi}\left(\frac{g\tau}{U}\right)^
{\frac{\pi}{2}\frac{U}{F_0}}
\quad (\mbox{DC, Large $U$})
\label{eq:GammadclargeU}
\end{eqnarray}
with the hopping parameter $\tau$ recovered. 
This result can be understood intuitively 
from a strong-coupling argument. 
After pair creation in the
Mott insulating ground state, the doublon must 
hop $\Delta_{\rm Mott}/F_0\simeq U/F_0$ sites
away from the accompanying hole to become on-shell. 
The amplitude decreases by a factor $\left(\frac{\tau}{U}\right)$
for each hopping, and thus, we are led to 
Eq.~(\ref{eq:GammadclargeU}).
In Fig.~\ref{fig:threshold}, we plot the $U$ dependence of the 
threshold (Schwinger limit).

\section{Keldysh crossover in AC fields}
\label{sec:AC}

\begin{figure}[tb]
\centering 
\includegraphics[width=8cm]{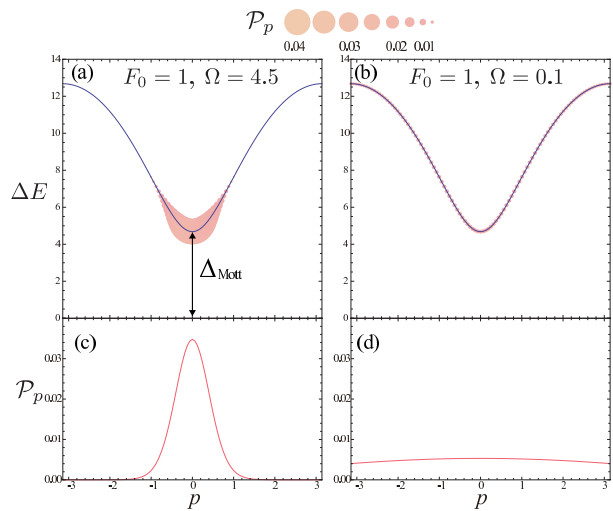}
\caption{(color online)
Tunneling probability of the dh-pair 
obtained for the $U=8$ Hubbard model in an AC fields. 
(a) and (c) correspond to  $F_0=1.0$ and $\Omega=4.5$,
respectively,
which are in the multi-photon regime, while 
(b) and (d) are for $F_0=1.0$, and $\;\Omega =0.1$,
respectively, which are
in the quantum tunneling regime. 
In (a) and (b), the tunneling probability is 
indicated by the size of the circle plotted on top
of the dh-pair spectrum. 
}
\label{fig:PpAC}
\end{figure}

\begin{figure*}[htb]
\centering 
\includegraphics[width=14cm]{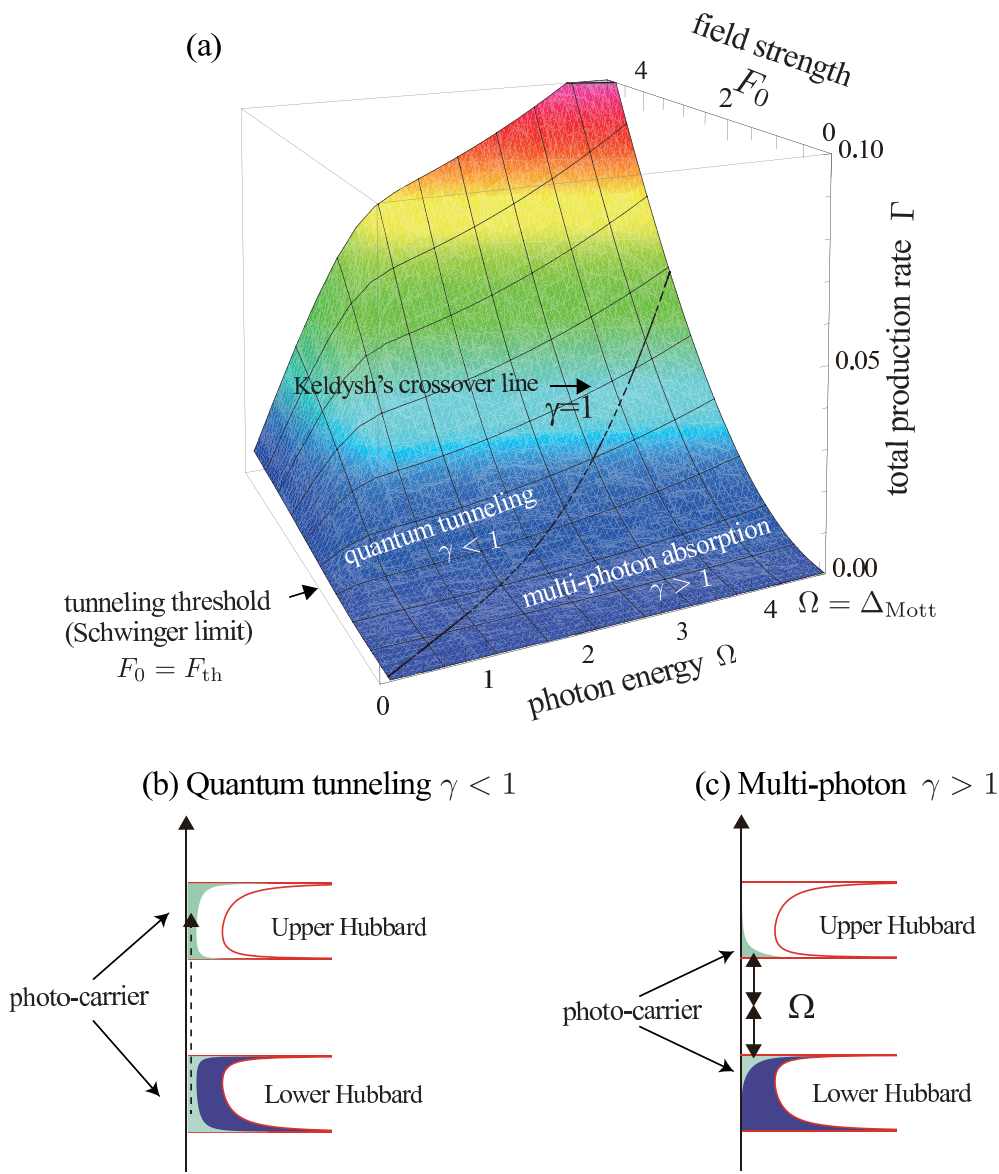}
\caption{(color online)
(a)
Total production rate of the $U=8$ Hubbard model in an AC field. 
The dashed line corresponds to 
Keldysh crossover $\gamma=1$, where the 
excitation mechanism changes from multi-photon absorption 
to quantum tunneling. 
Schwinger limit is $F_{\rm th}=1.668$.
(b), (c) Schematic pictures of the nonlinear excitation 
in the two regimes.
In the quantum tunneling regime (Fig.~(b)), the 
doublon-hole distribution becomes momentum independent, 
which means that the ``upper Hubbard band" becomes 
populated by photocarriers. 
}
\label{fig:GammaAC}
\end{figure*}
Next, we study a situation where a strong laser represented by
\begin{eqnarray}
F(t)=F_0 \sin\Omega t
\end{eqnarray}
($F_0$: field strength, $\Omega$: photon energy)
is applied to a Mott insulator. 
Experimentally, this  models photocarrier
injection, which is the initial process in the
photo induced insulator to metal phase transition
\cite{Iwai2003,Wall2011}. 
In standard photocarrier injection, 
the laser's photon energy $\Omega$
is set to the absorption peak, which is slightly 
above the Mott gap $\Delta_{\rm Mott}$. 
However, herein, we are interested in the nonlinear process
induced by subgap lasers, i.e., $\Omega<\Delta_{\rm Mott}$.

The tunneling probability $\mathcal{P}_p$
as well as production rate $\Gamma_p$
can be calculated using Eq.~(\ref{eq:LD2}) and Table \ref{table:Jacobian}. 
We note that the sign in the Jacobian
is determined so that $\mbox{Im}\mathcal{D}_{p1,2}\ge 0$ is satisfied.

An interesting feature of photocarriers
generated by nonlinear subgap lasers
is that {\it one can control the 
distribution of doublon-hole pairs by changing the photon
energy $\Omega$}. 
In Fig.~\ref{fig:PpAC}, we plot the momentum resolved tunneling probability 
for the $U=8$ Hubbard model. 
We notice that the distribution in the momentum space changes drastically when 
photon energy is changed. 
When $\Omega$ is large, the generated dh-pair is localized near 
the gap $\Delta E\sim \Delta_{\rm Mott}$. 
The peak becomes broader as the field strength $F_0$ becomes larger. 
On the other hand, when $\Omega$ is small, 
the dh-pair becomes uniformly distributed in the $p$ space. 
In the small $\Omega$ limit, we approach the DC field case,
where tunneling probability has no 
$p$-dependence, c.~f., Eq.~(\ref{eq:DC}).
In fact, the excitation mechanisms in the two regimes 
are different. 
For small $\Omega$ and large field strength, 
quantum tunneling is responsible for dh-pair creation.
On the other hand, when $\Omega$ is large, multi-photon absorption is the 
excitation mechanism. 
If we change photon energy and laser strength, 
there is a crossover between the two regimes, which is 
the Keldysh crossover \cite{Keldysh65} mentioned in the Introduction. 
We can directly see this from the 
analytical expression of the $p=0$ tunneling probability 
\begin{eqnarray}
\mathcal{P}_{p=0}\simeq\exp\left(-\frac{2\Delta_{\rm Mott}\gamma}
{\Omega}f(\gamma)\right)\qquad (\mbox{AC-fields}),\\
\to\left\{
	\begin{array}{cc}
\left(\frac{F_0\xi}{h\Omega}\right)^{2\frac{\Delta_{\rm Mott}}{\Omega}}
&\gamma\gg 1,\\
\exp\left(
-\frac{\pi}{2}\frac{\Delta_{\rm Mott}}{\xi F_0}
(1-\frac{\pi}{16}\gamma^2+\ldots)
\right)&\gamma\ll 1,
\end{array}
\right.\
\label{crossover}
\end{eqnarray}
This result is obtained with the help of the 
approximation in Eq.~(\ref{eq:ImE})
($h \sim 1.47$ and see footnote\cite{footnoteKE} for function $f$).
We note that this expression is 
identical to the QED result \cite{PopovJETP1972} with a redefinition
of Keldysh's adiabaticity parameter \cite{Keldysh65}
\begin{eqnarray}
\gamma=
\frac{\Omega}{\xi F_0}.
\label{eq;Keldysh}
\end{eqnarray}
The crossover is characterized by the 
Keldysh line defined by $\gamma=1$. 
In the multi-photon absorption regime 
[$\gamma\ll 1$; Fig.~\ref{fig:GammaAC}(c)],
tunneling probability has a power law dependence on 
field strength.  
The power $2\Delta_{\rm Mott}/\Omega$ 
is twice the number of absorbed photons. 
As stated above, photocarriers are generated near the
excitation gap. 

On the other hand, in the quantum tunneling regime
[$\gamma\gg 1$; Fig.~\ref{fig:GammaAC}(b)], 
tunneling probability shows a threshold behavior
with an exponential suppression. 
In this regime, the photocarriers are distributed almost equally 
in the momentum space. 
In Fig.~\ref{fig:GammaAC} (a), we plot the
total production rate as a function of photon energy and
field strength. The Keldysh crossover line is indicated by 
a dashed line. In the quantum tunneling regime, 
the production rate quickly increases as 
field strength exceeds tunneling threshold, 
which is $F_{\rm th}=1.668$ for the $U=8$ Hubbard model.

When we compare the present result with those of previous 
studies on quantum tunneling in AC field backgrounds such as
 atom ionization \cite{Keldysh65} and 
nonlinear QED\cite{Brezin1970,PopovSovJNP1974,PopovJETP1972},
we notice that the Keldysh crossover is quite universal and is
not limited to the Hubbard model. 
Previous studies were conducted on non-interacting 
systems where excitation occurs between 
single particle gaps, whereas 
in the present case, the system strongly interacts
and the origin of the excitation gap is a many-body effect.
The basic idea of the Keldysh crossover 
survives in many-body systems
and expressions such as tunneling threshold (Schwinger limit)
[Eq.~(\ref{eqn:Schwingerlimit})]
and the Keldysh's adiabaticity parameter (Eq.~(\ref{eq;Keldysh}))
are valid where the many-body features are renormalized
on the correlation length $\xi$. 
However, if we carefully consider 
the long time dynamics, differences between 
non-interacting systems and many-body systems can be observed. 
This is examined below in the next subsection.

\subsection{Comparison with numerical results}

\begin{figure}[htb]
\centering 
\includegraphics[width=7cm]{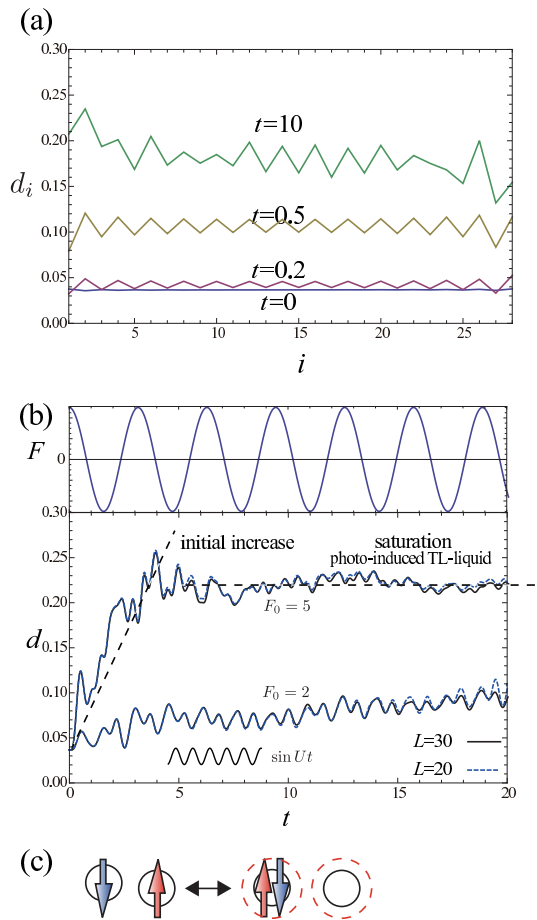}
\caption{(color online)
(a) Time evolution of doublon density in AC-fields
calculated by td-DMRG for the $U=8$ Hubbard model
($L=30,\;F_0=4,\;\Omega=2$).
(b) Time evolution of the averaged doublon density $d$. 
Results for $L=30$ and $L=20$ are plotted. 
Upper panel is the electric field $F(t)=F_0\cos\Omega t$.
Doublon density shows intial increase 
and then  saturation occurs. 
(c) Resonant oscillation between the ground state 
and the excited state with a neighboring doublon and hole pair. 
}
\label{fig:tddmrg}
\end{figure}

\begin{figure}[htb]
\centering 
\includegraphics[width=8.5cm]{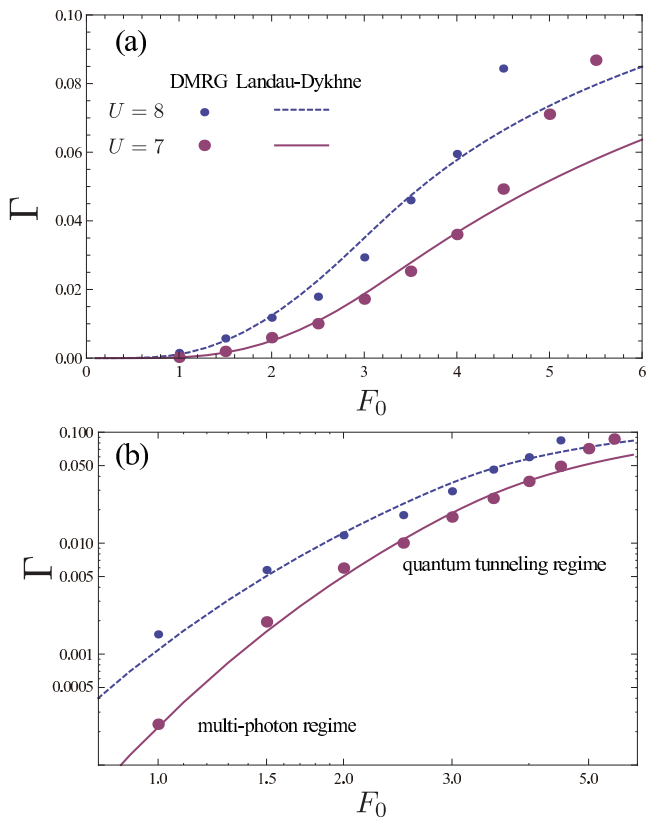}
\caption{(color online)
Total production rate for the $U=8,\;7$ 
Hubbard model with photon energy $\Omega=1$
plotted against field strength $F_0$ ((b) is the logarithmic plot).
The td-DMRG result is obtained using the fitting of
Eq.~(\ref{eq:DMRGfit}),
which is compared with the Landau-Dykhne result. 
Photon energy is $\Omega=1$
and system size is $L=30$. 
}
\label{fig:GammaDMRG}
\end{figure}

To examine the applicability
of the  Landau-Dykhne method, 
we compare its result with that of
td-DMRG \cite{SchollwoeckReview}. 
The 1D Hubbard model on an $L$-site 
open chain with the Hamiltonian 
\begin{eqnarray}
H(t)&=&-\tau\sum_{j,\sigma}(c^\dagger_{j+1\sigma}c_{j\sigma}+c^\dagger_{j\sigma}c_{j+1\sigma})
\label{eqn:HubbardHamiltonian2}\\
&&+U\sum_jn_{j\up}n_{j\dw}+F(t)\sum_jjn_j\nonumber.
\end{eqnarray}
is studied. This Hamiltonian is 
identical to the previous one 
[Eq.~(\ref{eqn:HubbardHamiltonian})] 
in the infinite size limit, and is related
through a gauge transformation. 
We start time evolution from the 
ground state $|0\ket$ obtained by the finite-size method
and apply the electric field for $t>0$. 
In the calculation, the DMRG Hilbert space is $m=200$, the time step is 0.01,
and the length of the chain is $L=20,\;30$.

There are several interesting features
in the time evolution of the doublon density
\begin{eqnarray}
d_i=\bra n_{i\up}n_{i\dw}\ket
\end{eqnarray}
as plotted in Fig.~\ref{fig:tddmrg}.
First, in addition to the overall increase, 
a density wave of doublons 
of the form $d_i(t)\sim d(t)+(-1)^i\delta d(t)$ occurs. 
We note that the electron density $n_i=\sum_\sigma \bra n_{i\sigma}\ket$
does not show such modulations. 
In addition, the finite-size effect is present in the calculation. 
This shows up in the
doublon density as a difference between the edge and bulk values. 
To eliminate the finite-size effect as well as the 
density wave from the analysis, we define the 
averaged doublon density $d(t)$ as an average within the 
central region of the chain. For example, for an $L=30$ system, 
the average is taken over the middle $10$ sites. 
The time evolution of the averaged doublon density 
is plotted in Fig.~\ref{fig:tddmrg}(b)
for $L=20$ and $L=30$, and we see that size dependence is not large
after averaging. 
The averaged doublon density shows a fast oscillation 
with a period of $2\pi/U$. 
One possible explanation
for the appearance of the density wave and time oscillation
is the locality of dh-pair production. 
The correlation length $\xi$ gives the typical size of the 
dh-excitation created by electric fields.
Because $\xi$ is short when $U$ is large, e.g.,
$\xi<1$ when $U>10$ [see Fig.~\ref{fig:xiDelta}(b)], 
most of the dh-excitation 
takes place between neighboring sites as in Fig.~\ref{fig:tddmrg}(c). 
The energy difference between the ground state
this state is $U$, which leads to temporal oscillation. 

The averaged doublon density plotted in  Fig.~\ref{fig:tddmrg}(b)
shows an increase and the speed of increase
becomes larger in stronger fields.
After the initial increase, we notice that saturation takes place
although the AC field is still present. 
If the field strength is sufficiently large, the long-term 
driven state is a metallic state called the 
``photo-induced Tomonaga-Luttinger-like liquid"\cite{OkaPhotoLuttinger}. 
In this state, it was numerically shown by calculating 
correlation functions that spin-charge separation takes place.
However, in this study, we restrict ourselves to nonlinear 
dh-pair creation, which is responsible for the initial increase in
doublon density. 
The production rate of the dh-pairs is obtained from numerical data
by employing the fitting 
\begin{eqnarray}
d(t)=d_0+a(\tanh(b t)+1)
\end{eqnarray}
and production rate is identified as the initial slope
\begin{equation}
\Gamma_{\rm DMRG}=ab.
\label{eq:DMRGfit}
\end{equation}
In Fig.\ref{fig:GammaDMRG} we plot the total production rate
obtained by td-DMRG and compare it with the analytical result calculated
from the Landau-Dykhne formula, 
i.e., Eq.~(\ref{eq:totalproductionrate}). 
Although the fitting is difficult owing to the 
fast oscillation of doublon density, the results 
seem to agree quite well. 
The numerical result agrees with the 
Landau-Dykhne result in the weak field 
regime up to $F_0<4$. 
The deviation at large $F_0$ is expected because the 
Landau-Dykhne formula is known to work only near the adiabatic limit, i.e., 
$\mathcal{P}\ll 1$. 
In the log-log plot, we clearly see the crossover from the 
weak field multi-photon absorption regime, with a power law behavior,
to the quantum tunneling regime.

\begin{figure*}[thb]
\centering 
\includegraphics[width=14cm]{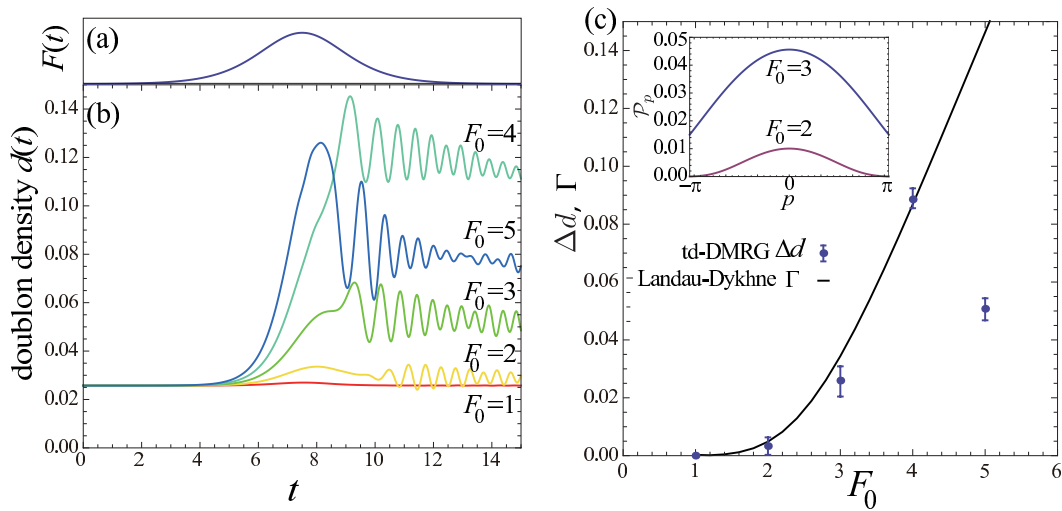}
\caption{(color online)
(a) Shape of pulse field with duration $\sigma=2$.
(b) Time evolution of the doublon density $d(t)$
for the $U=10$ Hubbard model obtained by td-DMRG.
(c) Increase in the doublon density $\Delta d$ obtained by td-DMRG
and the total production $\Gamma$ obtained 
by the Landau-Dykhne method. 
See main text for the error bar. 
Inset: $p$-resolved distribution function of doublon-hole 
pairs obtained by the Landau-Dykhne method.
}
\label{fig:pulse}
\end{figure*}

\subsection{Single pulse} 
\label{sec:singlepulse}
Using the Landau-Dykhne method, we can 
extend the analysis to the case of a single pulse. 
There are two motivations.
First, recent ultra fast pulse lasers are  
becoming so short that a single pulse excitation is now experimentally 
realizable. Second, we wish to further examine the validity of the 
Landau-Dykhne method from a theoretical perspective.
In this regard, a single pulse is 
ideal because problems such as saturation and 
driven steady states do not occur, as is the case
with a continuous AC field. 
The pulse we study here is 
\begin{eqnarray}
F=F_0\cosh^{-2}(t/\sigma)
\end{eqnarray}
and the Jacobian needed for the calculation
is given in Table \ref{table:Jacobian}.
We set the duration to $\sigma=2$.

In Fig.~\ref{fig:pulse}(b), we plot the 
time evolution of doublon density 
for a 1D $U=10$ Hubbard model on an open $L=50$ chain
calculated by td-DMRG.
Doublon density initially increases and
then oscillates with a period 
roughly given by 
$T=2\pi/U$.
This oscillation is similar to the AC field 
case seen in Fig.~\ref{fig:tddmrg}(b).
Increase in the doublon density $\Delta d$
is plotted in Fig.~\ref{fig:pulse}(c),
where the error bars are determined by the 
oscillation width near $t=15$.
At $F_0<4$, $\Delta d$ shows a 
monotonic increase, whereas it shows an 
irregular decrease at $F_0=5$.

Let us compare the numerical result with
Landau-Dykhne results.
In the inset of Fig.~\ref{fig:pulse}(c), 
the momentum resolved tunneling probability 
$\mathcal{P}_p$ is plotted.  
The distribution has a peak at $p=0$ 
and broadens as the field becomes stronger. 
Eventually, as seen in the $F_0=3$ data, 
high-energy dh-pairs with momentum $p\sim \pm \pi$
become excited as well. 
We compare the total production rate $\Gamma$ obtained 
from the Landau-Dykhne approach
with $\Delta d$.
Comparison with td-DMRG revealthat the Landau-Dykhne method
gives the correct threshold behavior 
and is reliable even for relatively strong electric fields.
However, the Landau-Dykhne method does not capture the 
irregular drop at $F_0=5$ and we detect 
a small deviation at $F_0=3$. 
As mentioned above, the Landau-Dykhne method 
ignores the interference effect due to  multiple quantum 
tunneling\cite{Vasilev2004}, 
which becomes important as 
pair annihilation of doublons and holes become activated. 
In fact, when the effect of interference is 
strong, quantum tunneling can be suppressed by 
dynamical localization in the energy space \cite{Oka2004a}.
Such an effect is expected to become important when the field is strong,
and we think that it explains the differences between 
the numerical and Landau-Dykhne approach results.

\section{Experimental feasibility}
\label{sec:exp}
In this section, we discuss the experimental 
feasibility of the nonlinear doublon excitations
predicted in this study. 
Candidates of physical systems range from 
fermionic cold atoms to solid state crystals. 
For example, if we use fermionic cold atoms in 
an optical lattice, 
the present theory can be experimentally verified 
by inducing a time-dependent tilt of the lattice. 
This mimics the effect of an electric field and 
a direct measurement of the doublon density
is possible as well \cite{StrohmainerEsslinger10,GreifEssslinger2011}.
In solid states, we must estimate the 
threshold field strengths of candidate materials 
and compare them with the peak field strengths of present day lasers. 
In addition, because doublon density is not a directly
observable quantity
in solids, we must seek alternative detection methods.
We discuss these issues below.

\subsection{Threshold of 1D Mott insulators}

\begin{table*}[tbh]
\centering 
\begin{tabular}{|c|ccc|ccc|}
\hline
&$\tau$(eV)&$U$(eV)&$a$($\AA$)&
$\Delta_{\rm Mott}$(eV)&$\xi$($a$)&$E_{\rm th}(\mbox{MV}/\mbox{cm})$\\\hline
ET-F$_2$TCNQ&0.1&1&10&0.7&1.1&3\\
$[$Ni(cnxn)$_2$Br$]$Br$_2$&0.22&2.4&5&1.6&1.0&16\\
Sr$_2$CuO$_3$&0.52&3.1&4&1.5&2.1&9\\
\hline
\end{tabular}
\caption{
Mott gap $\Delta_{\rm Mott}$, correlation length $\xi$, and 
tunneling threshold (Schwinger limit) $F_{\rm th}=eaE_{\rm th}$
calculated by the Landau-Dykhne method with the Bethe ansatz.
Material parameters ($\tau$: hopping, $U$: onsite repulsion,
$a$: lattice constant)
are obtained from Ref.~\cite{Wall2011}~(ET-F$_2$TCN)
and Ref.~\cite{Tomita2001}~($[$Ni(cnxn)$_2$Br$]$Br$_2$, Sr$_2$CuO$_3$).}
\label{table:Fsch}
\end{table*}

Pump-probe experiments with a terahertz (THz) laser 
are ideal setups to verify the Keldysh crossover
described in section \ref{sec:AC}.
Current THz pulse lasers 
can be as strong as $1\;\mbox{MV/cm}$ 
\cite{HiroriAPL2011,Watanabe2011} and the
typical photon energy is $\Omega=4\;\mbox{meV}$ 
corresponding to 1 THz.

On the material side, the candidate 
1D Mott insulator ranges from 
organic crystals to cuprates. 
The tunneling thresholds of several materials estimated by
the Landau-Dykhne + Bethe ansatz
method are shown in table~\ref{table:Fsch}.
In the list, the material with the 
smallest threshold is ET-F$_2$TCNQ (ET-salt).
Let us discuss the possibility of generating 
dh-pairs in this material with a THz laser. 
The tunneling threshold (Schwinger limit) for 
the material is $E_{\rm th}=3\;\mbox{MV/cm}$. 
When using a laser with peak strength
$E_{\rm laser}=1\;\mbox{MV/cm}$ \cite{HiroriAPL2011,Watanabe2011}, 
we are still below the threshold,
and the tunneling probability is 
\begin{eqnarray}
\mathcal{P}&=&\exp\left(-\pi\frac{E_{\rm th}}{E_{\rm laser}}\right)\nonumber\\
&\sim&
8\times 10^{-5}\quad 
(\mbox{for ET-salt}).
\end{eqnarray}
This is too small 
to trigger photo-induced metallization, and therefore,
a stronger light source is needed.  
Quite recently, amplification of
laser field strength using a metamaterial structure  
has been proposed\cite{Hwang2012APS,LiuNat2012}. 
It is reported that peak strength can be as large as
$
E_{\rm metamaterial}=$4 MV/cm   \cite{LiuNat2012}.
If this technique can be applied, 
tunneling probability becomes
\begin{eqnarray}
\mathcal{P}&=&\exp\left(-\pi\frac{E_{\rm th}}{E_{\rm metamaterial}}\right)\nonumber\\
&\sim&
0.1\quad 
(\mbox{for ET-salt}).
\end{eqnarray}
This means than one can perform a $\sim$10\% photodoping with a single
pulse. This value is large enough that one can 
trigger a photo-induced phase transition.

\subsection{Angle resolved photoemission spectroscopy 
of the upper Hubbard band in the quantum tunneling regime}
As stated in Section \ref{sec:AC} [see Fig.~\ref{fig:GammaAC}(b),~(c)],
the distribution of the produced dh-pairs changes drastically 
when the laser is shifted from the multi-photon absorption regime 
to the quantum tunneling regime. 
In the quantum tunneling regime, high energy dh-pairs are generated.
This means that the entire ``upper Hubbard band" becomes 
populated by photocarriers. 
Thus, using a strong THz laser as the pump, 
it is possible to study the band structure of the 
upper Hubbard band with a real time 
angle resolved photoemission spectroscopy technique. 
A necessary condition is that the Keldysh parameter (Eq.~(\ref{eq;Keldysh}))
\begin{eqnarray}
\gamma=
\frac{\Omega}{\xi F_0}.
\label{eq;Keldysh2}
\end{eqnarray}
is well below unity. 
From Table~\ref{table:Fsch}, the Keldysh crossover field strength for a
1 THz laser ($\Omega=4\;\mbox{meV}$) is
\begin{eqnarray}
E_{\rm crossover}&=&\frac{\Omega}{a\xi}\nonumber\\
&\sim&
4\times 10^{-2}\;\mbox{MV/cm}\quad 
(\mbox{for ET-salt}),
\end{eqnarray}
and the pump laser must exceed this strength. 
With stronger fields, more photocarriers are excited and 
the measurement is expected to 
be more feasible.

\subsection{Dielectric breakdown and nonlinear transport}
\label{sec:breakdown}

\begin{figure}[tbh]
\centering 
\includegraphics[width=5cm]{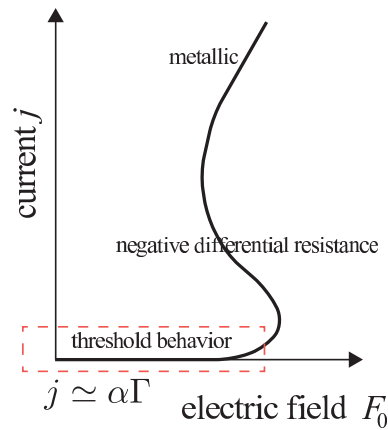}
\caption{
(color online)
Typical $IV$-characteristics of strongly correlated insulators.
We have a threshold behavior, negative differential resistance
and a transition to a metallic state.  
The present theory is only applicable to 
explain the threshold behavior
in the small current regime shown in the dashed box.
}
\label{fig:iv}
\end{figure}
In nonlinear transport experiments, a
threshold behavior in the $IV$-characteristics
is found in many correlated insulators. 
Materials range from Mott insulators \cite{tag},
charge-ordered systems \cite{Moriheating}, and
materials showing a neutral-ionic transition\cite{Tokura88}. 
In Fig.\ref{fig:iv}, we plot typical $IV$-characteristics. 
In many cases, a threshold behavior as well as 
a region with negative differential resistance 
is present. 
Let us make a comment on the threshold behavior appearing
in the small current regime, i.e.,
dielectric breakdown [region inside the dashed line in Fig.\ref{fig:iv}].

Although doublon density 
is not a measurable quantity,
we can relate the dh-pair production rate to current itself. 
Theoretical studies \cite{PhysRevLett.105.146404,Tanaka2011}
suggests that, in nonlinear transport 
of Mott insulators, current has two major contributions
\begin{eqnarray}
J=\sigma F_0+\alpha \Gamma.
\label{eq:current}
\end{eqnarray}
The first term is the standard
linear response due to thermal carriers
with a temperature dependence
$\sigma_{\rm thermal}\propto e^{-\Delta_{\rm Mott}/T}$. 
The second term is proportional to the dh-pair production rate
with a nonperturbative threshold form
\begin{eqnarray}
\Gamma= \frac{1}{2\pi}\exp\left(-\pi\frac{F_{\rm th}}{F_0}\right)F_0.
\end{eqnarray}
$\alpha$ is a nonuniversal proportionality constant
which depends on the coupling to the electrode and other factors. 

One may think that photocarriers 
induced by dh-pair production may contribute 
to linear response, i.e., the term $\sigma F_0$. 
This is not true in dielectric breakdown
occuring in the quantum tunneling regime. 
The reason is because the dh-pairs are in 
an ``infinite temperature state" \cite{Philippprivate}. 
As shown in Section \ref{sec:DC},
the distribution of dh-pairs 
has no momentum-, and thus energy-, dependences. 
This corresponds to an infinite temperature state, i.e., 
$e^{-E/k_BT}$ with $T\to \infty$.
The conductivity of an infinite temperature state is zero,
and therefore, there are no linear reponse contribution in the 
quantum tunneling regime. 
Instead, the current is dominated by 
the second term in Eq.~(\ref{eq:current}) which is 
proportional to the dh-pair production rate. 
The doublons and holes are pair created, 
separated from each other by the electric field, 
and are measured as current when they are
absorbed by electrodes. 
This feature is consistent with numerical results 
obtained in a static system 
coupled to electrodes \cite{Tanaka2011}.

After dielectric breakdown, the $IV$-characteristics show
interesting nonlinear behaviors. 
Although this is far beyond the applicability of the present 
theory, let us consider existing literatures. 
If voltage drop is measured as a function of 
current, there is a regime where
negative differential resistance is realized \cite{Tokura88,tag,Moriheating}. 
The origin of negative differential resistance is not fully understood yet. 
It was pointed out, with a careful comparison with 
experimental data, that the 
temperature increase of the sample
due to Joule heating can explain it \cite{Moriheating}. 
They suggest that negative differential resistance 
occurs when the temperature dependence of conductivity is large. 
A related theoretical paper 
explained negative differential resistance
 in disordered films via the heating mechanism
\cite{Altshuleroverheating09}. 
A more dramatic proposal is based on
nonequilibrium first-order phase transition proposed by 
Ajisaka {\it et.~al.} \cite{Ajisaka2009}, 
where negative differential resistance is 
explained through a phase bi-stability. 
Negative differential resistance was also found in a model in high energy physics, namely the 
supersymmetric QCD in the large $N$ limit \cite{Nakumura2010}. 
A microscopic understanding of the nonlinear transport 
properties of correlated systems from a universal viewpoint 
is an interesting callenge.

\subsection{Optical sum rule}
From  Eq.~(\ref{eq:current}), 
we can derive an interesting relationship between the 
optical sum rule and the doublon production rate
in DC-electric fields. 
Here, we consider the low temperature
case where contributions from thermal carriers are negligible, 
and the field strength is below the 
threshold [region inside dashed line in Figure~\ref{fig:iv}].

The optical sum rule is commonly used by experimentalists 
as a means to ``measure" carrier density 
from the absorption spectrum\cite{Iwai2003}. 
The optical sum rule for the Hubbard model
(e.g., \cite{JeckelmannPRL00})
states that the 
frequency integral of the absorption spectrum,
which we call $N_{\rm eff}$ following Ref.~~\cite{Iwai2003},
is related to kinetic energy as follows:
\begin{eqnarray}
N_{\rm eff}&\equiv&\int_0^\infty \frac{d\omega}{\pi}\sigma_1(\omega)=-\frac{1}{2L}\overline{K},
\end{eqnarray}
Here, $\overline{K}$ is the time average of the
 expectation value of the kinetic term 
in the Hamiltonian, i.e.,
\begin{eqnarray}
K(t)=\bra \Psi(t)|-
\tau\sum_{j,\sigma}(e^{i\Phi}c^\dagger_{j+1\sigma}c_{j\sigma}+e^{-i\Phi}c^\dagger_{j\sigma}c_{j+1\sigma})|\Psi(t)\ket.\nonumber\\
\end{eqnarray}
Below the threshold, dh-pair creation is a rare event and 
current is very small. 
In this regime, energy dissipation 
by external degrees of freedom, e.g., phonons, is negligible
and we can use Joule's relation 
\begin{eqnarray}
\frac{d}{dt}\bra H(t)\ket=LJF_0
\end{eqnarray}
to relate change in energy
\begin{eqnarray}
\bra H(t)\ket=K(t)+ULd(t)
\end{eqnarray}
to the current $J$. 
Using Eq.~(\ref{eq:current}) 
as well as $\frac{d}{dt}d(t)=\Gamma$, 
we obtain 
\begin{eqnarray}
\frac{d}{dt}N_{\rm eff}=(U-\alpha F_0)\Gamma.
\label{eq:dNdt}
\end{eqnarray}
This formula states that, in static electric fields, 
optical sum $N_{\rm eff}$
increases linearly in time and the 
speed is proportional to the dh-pair production rate. 
This relationship opens a way to measure 
doubon increase by optical experiments.

\section{ Conclusion}
We developed an analytical 
theory for nonlinear pair excitations of doublons and holes
in a 1D Mott insulator subject to DC, AC and 
pulse electric fields. 
The theory is based on the Landau-Dykhne method 
combined with the Bethe ansatz.
In an AC-field, the theory predicts a crossover between 
multi-photon absorption and quantum tunneling 
when the strength and photon energy of the field changes. 
Comparison with numerical results by td-DMRG
shows that the analytical theory is reliable up 
to moderate field strength.

There are several limitations in our theory. 
Perhaps one of the most important open issues is the
treatment of temperature effects. 
It is unclear if the tunneling probability 
has a direct temperature dependence. 
Numerical results obtained by nonequilibrium 
DMFT suggest no or very small temperature dependence
in the quantum tunneling regime \cite{PhysRevLett.105.146404},
whereas strong temperature
dependence of the threshold is seen in a dielectric breakdown experiment
\cite{tag}. 
Impact ionization, an avalanche like cascade growth of carriers  
due to field induced acceleration, may be important
in understanding these experiments. 
The origin of negative differential resistance and the 
properties of the nonequilibrium steady 
state is another important 
open problem (Section~\ref{sec:breakdown}). 
We think that trying to answer these problems will lead 
to important innovations in nonequilibrium manybody physics.

We acknowledge Philipp Werner, Kunio Ishida, David Pekker, Rajdeep
Sensarma, Li Gao, Stuart Parkin, Gerald Dunne and Eugene Demler for valuable
discussions. 
TO acknowledges support from 
Grant-in-Aid for Young Scientists (B), CUA and ITAMP.

{\bf Note added:}
After the submission of the initial version of the manuscript, 
Lenar\v{c}i\v{c} and Prelov\v{s}ek 
published an interesting paper\cite{Lenarcic12}.
They studied the 
dielectric breakdown in a spin polarized 
Mott insulator
and found that the threshold has a 
$F_{\rm th}\propto \Delta^{3/2}$ dependence, 
 which is different from the conventional Landau-Zener form. 
They pointed out that the origin of this difference
comes from excitation dispersion. 
In the half-filled case, the dispersion
is relativistic $\omega_k\propto (k^2+\kappa^2)^{1/2}$,
i.e., Eq.~(\ref{eq:ImE}), whereas
it is parabolic $\omega_k\propto k^2+\kappa^2$ in the 
spin polarized model. 
Using Eq.~(\ref{eq:Schwinger}), 
we can recover the $F_{\rm th}\propto \Delta^{3/2}$ behavior
in the spin polarized case. 
In another article, a variant of Eq.~(\ref{eq:Schwinger}) was used to 
study excitations in the 
attractive Hubbard model\cite{Uchino12}. 
These examples show the wide applicability of the
Landau-Dykhne approach in many-body problems.


\begin{thebibliography}{30}
\expandafter\ifx\csname natexlab\endcsname\relax\def\natexlab#1{#1}\fi
\expandafter\ifx\csname bibnamefont\endcsname\relax
  \def\bibnamefont#1{#1}\fi
\expandafter\ifx\csname bibfnamefont\endcsname\relax
  \def\bibfnamefont#1{#1}\fi
\expandafter\ifx\csname citenamefont\endcsname\relax
  \def\citenamefont#1{#1}\fi
\expandafter\ifx\csname url\endcsname\relax
  \def\url#1{\texttt{#1}}\fi
\expandafter\ifx\csname urlprefix\endcsname\relax\def\urlprefix{URL }\fi
\providecommand{\bibinfo}[2]{#2}
\providecommand{\eprint}[2][]{\url{#2}}

\bibitem[{\citenamefont{{S. Iwai M. Ono A. Maeda H. Matsuzaki H. Kishida H.
  Okamoto and Y. Tokura}}(2003)}]{Iwai2003}
\bibinfo{author}{\bibnamefont{{S. Iwai, M. Ono, A. Maeda, H. Matsuzaki, H. Kishida,  H. Okamoto, and Y. Tokura
}}}, \bibinfo{journal}{Phys. Rev. Lett.}
  \textbf{\bibinfo{volume}{91}}, \bibinfo{pages}{057401}
  (\bibinfo{year}{2003}).



\bibitem[{\citenamefont{{S. Wall, D. Brida, S. R. Clark, H. P. Ehrke, D.
  Jaksch, A. Ardavan, S. Bonora, H. Uemura, Y. Takahashi, T. Hasegawa, H.
  Okamoto, G. Cerullo, and A. Cavalleri}}(2011)}]{Wall2011}
\bibinfo{author}{\bibnamefont{{S. Wall D. Brida, S. R. Clark, H. P. Ehrke, D.  Jaksch, A. Ardavan, S. Bonora, H. Uemura, Y. Takahashi, T. Hasegawa, H.  Okamoto, G. Cerullo, and A. Cavalleri
}}}, \bibinfo{journal}{Nat. Phys.}
  \textbf{\bibinfo{volume}{7}}, \bibinfo{pages}{114} (\bibinfo{year}{2011}).

\bibitem{tag}
Y.~Taguchi,~T. Matsumoto, and Y. Tokura, Phys. Rev. B {\bf 62},  7015  (2000).

\bibitem{Moriheating}
T.~Mori, T.~Ozawa, Y.~Bando, T.~Kawamoto, S.~Niizeki, H.~Mori, 
and I.~Terasaki, Phys. Rev. B {\bf 79}, 115108 (2009),
T.~S.~Inada, I.~Terasaki, H.~Mori, and T.~Mori,
Phys. Rev. B {\bf 79}, 165102 (2009).

\bibitem{Tokura88}
Y.~Tokura, H.~Okamoto, T.~Koda,
T.~Mitani, and G.~Saito,
Phys. Rev. B {\bf 38}, 2215 (1988).


\bibitem[{\citenamefont{{S. Watanabe, N. Minami, R.
  Shimano}}(2011)}]{Watanabe2011}
\bibinfo{author}{\bibnamefont{{S. Watanabe, N. Minami, and R. Shimano}}},
  \bibinfo{journal}{Opt. Express} \textbf{\bibinfo{volume}{19}},
  \bibinfo{pages}{1528} (\bibinfo{year}{2011}).

\bibitem{HiroriAPL2011}
H. Hirori, A. Doi, F. Blanchard, and K. Tanaka,
Appl. Phys. Lett. {\bf 98} 091106 (2011).

\bibitem{LiuNat2012}
Mengkun Liu, Harold Y. Hwang, Hu Tao, Andrew C. Strikwerda1, 
Kebin Fan, George R. Keiser, Aaron J. Sternbach,
Kevin G. West, Salinporn Kittiwatanakul, Jiwei Lu, 
Stuart A.Wolf, Fiorenzo G. Omenetto, Xin Zhang, Keith A. Nelson,
and  Richard D. Averitt,
Nature 487, 345 (2012).

\bibitem[{\citenamefont{{M. Greiner O. Mandel T. Esslinger T. W. H\"ansch, I.l
  Bloch }}(2002)}]{Greiner2002}
\bibinfo{author}{\bibnamefont{{M. Greiner O. Mandel T. Esslinger T. W.
  H\"ansch, and I. Bloch }}}, \bibinfo{journal}{Nature}
  \textbf{\bibinfo{volume}{415}}, \bibinfo{pages}{39} (\bibinfo{year}{2002}).

 \bibitem{StrohmainerEsslinger10}
  N. Strohmaier, D. Greif, R. J\"ordens, L. Tarruell, H. Moritz, T. Esslinger, R. Sensarma, D. Pekker, E. Altman, and E. Demler, 
  \newblock Phys. Rev. Lett. {\bf 104}, 080401 (2010).

  \bibitem{GreifEssslinger2011}
D. Greif, L. Tarruell, T. Uehlinger, R. J\"ordens, and T. Esslinger 
  \newblock Phys. Rev. Lett. {\bf 106}, 145302 (2011).


\bibitem{Kollath06prl}
C.~Kollath, A.~Iucci, T.~Giamarchi, W.~Hofstetter, and U.~Schollw\"ock, 
Phys. Rev. Lett. {\bf 97}, 050402 (2006).


\bibitem[{\citenamefont{{T. Oka, R. Arita, and H. Aoki}}(2003)}]{Oka2003}
\bibinfo{author}{\bibnamefont{{T. Oka, R. Arita, and H. Aoki}}},
  \bibinfo{journal}{Phys. Rev. Lett.} \textbf{\bibinfo{volume}{91}},
  \bibinfo{pages}{066406} (\bibinfo{year}{2003}).

\bibitem[{\citenamefont{{T. Oka and H. Aoki}}(2005)}]{Oka2005a}
\bibinfo{author}{\bibnamefont{{T. Oka and H. Aoki}}}, \bibinfo{journal}{Phys.
  Rev. Lett.} \textbf{\bibinfo{volume}{95}}, \bibinfo{pages}{137601}
  (\bibinfo{year}{2005}).


\bibitem[{\citenamefont{{T.~Oka and H.~Aoki}}(2010)}]{OkaSchwinger10}
\bibinfo{author}{\bibnamefont{{T.~Oka and H.~Aoki}}}, \bibinfo{journal}{Phys.
  Rev. B} \textbf{\bibinfo{volume}{81}}, \bibinfo{pages}{033103}
  (\bibinfo{year}{2010}).

\bibitem{Oka2005b}
T.~Oka, N.~Konno, R.~Arita, and H.~Aoki,
Phys.~Rev.~Lett. {\bf 94} 100602 (2005).

\bibitem[{\citenamefont{Eckstein et~al.}(2010)\citenamefont{Eckstein, Oka, and
  Werner}}]{PhysRevLett.105.146404}
\bibinfo{author}{\bibfnamefont{M.}~\bibnamefont{Eckstein}},
  \bibinfo{author}{\bibfnamefont{T.}~\bibnamefont{Oka}}, \bibnamefont{and}
  \bibinfo{author}{\bibfnamefont{P.}~\bibnamefont{Werner}},
  \bibinfo{journal}{Phys. Rev. Lett.} \textbf{\bibinfo{volume}{105}},
  \bibinfo{pages}{146404} (\bibinfo{year}{2010}).


\bibitem{Eckstein2011}
M.~Eckstein and P.~Werner,
Phys. Rev. B {\bf 84}, 035122 (2011).




\bibitem{Tanaka2011}
Y.~Tanaka, and K.~Yonemitsu,
Phys. Rev. B {\bf 83}, 085113 (2011).

\bibitem[{\citenamefont{{F.~Heidrich-Meisner {\it et
  al}}}(2010)}]{Heidrich-Meisner10}
\bibinfo{author}{\bibnamefont{{F.~Heidrich-Meisner,
I. Gonz\'alez, K. A. Al-Hassanieh, A. E. Feiguin, M. J. Rozenberg, and E. Dagotto }}},
  \bibinfo{journal}{Phy. Rev. B} \textbf{\bibinfo{volume}{82}},
  \bibinfo{pages}{205110} (\bibinfo{year}{2010}).


\bibitem[{\citenamefont{{A. Takahashi, H. Itoh and M.
  Aihara}}(2008)}]{Takahashi08B}
\bibinfo{author}{\bibnamefont{{A. Takahashi, H. Itoh and M. Aihara}}},
  \bibinfo{journal}{Phys. Rev. B} \textbf{\bibinfo{volume}{77}},
  \bibinfo{pages}{205105} (\bibinfo{year}{2008}).


\bibitem{OkaPhotoLuttinger}
T. Oka and H. Aoki, Phys. Rev. B {\bf 78}, 241104 (R) (2008).


\bibitem[{\citenamefont{{T. Oka, N. Konno, R. Arita, and H.
  Aoki}}(2005)}]{Oka2004a}
\bibinfo{author}{\bibnamefont{{T. Oka, N. Konno, R. Arita, and H. Aoki}}},
  \bibinfo{journal}{Phys. Rev. Lett.} \textbf{\bibinfo{volume}{94}},
  \bibinfo{pages}{100602} (\bibinfo{year}{2005}).

\bibitem{SachdevElectricField02}
S. Sachdev, K. Sengupta, and S. M. Girvin, 
Phys. Rev. B {\bf 66}, 075128 (2002).

\bibitem{Ajisaka2009}
S.~Ajisaka, H.~Nishimura, S.~Tasaki, and I.~Terasaki, 
Prog. Theor. Phys. {\bf 121} 1289 (2009).

\bibitem[{\citenamefont{Nakamura}(2010)}]{Nakumura2010}
\bibinfo{author}{\bibfnamefont{S.}~\bibnamefont{Nakamura}},
  \bibinfo{journal}{Prog. Theor. Phys.} \textbf{\bibinfo{volume}{124}},
  \bibinfo{pages}{1105} (\bibinfo{year}{2010}).


\bibitem[{\citenamefont{{M. Imada, A. Fujimori, and Y.
  Tokura}}(1998)}]{Imada1998}
\bibinfo{author}{\bibnamefont{{M. Imada, A. Fujimori, and Y. Tokura}}},
  \bibinfo{journal}{Rev. Mod. Phys.} \textbf{\bibinfo{volume}{70}},
  \bibinfo{pages}{1039} (\bibinfo{year}{1998}).


\bibitem[{\citenamefont{{A. M. Dykhne}}(1962)}]{Dykhne1962}
\bibinfo{author}{\bibnamefont{{A. M. Dykhne}}}, \bibinfo{journal}{Sov. Phys.
  JETP} \textbf{\bibinfo{volume}{14}}, \bibinfo{pages}{941}
  (\bibinfo{year}{1962}).

\bibitem[{\citenamefont{{L. Landau and E. Lifshitz}}(1981)}]{LandauLifshitzQM}
\bibinfo{author}{\bibnamefont{{L. Landau and E. Lifshitz}}},
  \emph{\bibinfo{title}{Quantum Mechanics, Vol. 3}}
  (\bibinfo{publisher}{Butterworth-Heinemann}, \bibinfo{year}{1981}).


\bibitem{DavisPechukas1976}
J.~P.~Davis and P.~Pechukas,
J. Chem. Phys. {\bf 64}, 3129 (1976).


\bibitem[{\citenamefont{Keldysh}(1965)}]{Keldysh65}
\bibinfo{author}{\bibfnamefont{L.}~\bibnamefont{Keldysh}},
  \bibinfo{journal}{JETP} \textbf{\bibinfo{volume}{20}}, \bibinfo{pages}{1307}
  (\bibinfo{year}{1965}).


 \bibitem{DeloneKrainovBook}
  N.B. Delone, and V.P. Krainov, {\it Multiphoton Processes in Atoms},
  (Springer, 2000).



\bibitem{Wilkinson2000}
M.~Wilkinson and M.~A.~Morgan,
Phys. Rev. A {\bf 61}, 062104 (2000).



\bibitem[{\citenamefont{Heisenberg and Euler}(1936)}]{Heisenberg1936}
\bibinfo{author}{\bibfnamefont{W.}~\bibnamefont{Heisenberg}} \bibnamefont{and}
  \bibinfo{author}{\bibfnamefont{H.}~\bibnamefont{Euler}},
  \bibinfo{journal}{Z.Physik} \textbf{\bibinfo{volume}{98}},
  \bibinfo{pages}{714} (\bibinfo{year}{1936}).

\bibitem[{\citenamefont{Schwinger}(1951)}]{Schwinger1951}
\bibinfo{author}{\bibfnamefont{J.}~\bibnamefont{Schwinger}},
  \bibinfo{journal}{Phys. Rev.} \textbf{\bibinfo{volume}{82}},
  \bibinfo{pages}{664} (\bibinfo{year}{1951}).

\bibitem[{\citenamefont{Popov}(1974)}]{PopovSovJNP1974}
\bibinfo{author}{\bibfnamefont{V.~S.} \bibnamefont{Popov}},
  \bibinfo{journal}{Sov. J. Nucl. Phys.} \textbf{\bibinfo{volume}{19}},
  \bibinfo{pages}{584} (\bibinfo{year}{1974}).

\bibitem[{\citenamefont{Popov}(1972)}]{PopovJETP1972}
\bibinfo{author}{\bibfnamefont{V.~S.} \bibnamefont{Popov}},
  \bibinfo{journal}{JETP} \textbf{\bibinfo{volume}{34}}, \bibinfo{pages}{709}
  (\bibinfo{year}{1972}).

\bibitem[{\citenamefont{Brezin and Itzykson}(1970)}]{Brezin1970}
\bibinfo{author}{\bibfnamefont{E.}~\bibnamefont{Brezin}} \bibnamefont{and}
  \bibinfo{author}{\bibfnamefont{C.}~\bibnamefont{Itzykson}},
  \bibinfo{journal}{Phys. Rev. D} \textbf{\bibinfo{volume}{2}},
  \bibinfo{pages}{1191} (\bibinfo{year}{1970}).


\bibitem{Popovreview}
V.~S.~Popov, Physcs-Uspekhi, {\bf 47} 855 (2004).

\bibitem{DumluDunne10}
C.~K.~Dumlu and G.~V.~Dunne, Phys.~Rev.~Lett. {\bf 104}
250402 (2010). 

\bibitem{Dunne04review}
G.~V.~Dunne, ``New Strong-Field QED Effects at ELI: Nonperturbative Vacuum Pair Production", arXiv:0812.3163v2,
 ``Heisenberg-Euler Effective Lagrangians : Basics and Extensions", arXiv:hep-th/0406216. 

\bibitem{Okareview}
T.~Oka and H.~Aoki, "Nonequilibrium Quantum Breakdown in a Strongly Correlated Electron System" in `` Quantum and Semi-classical Percolation and Breakdown in Disordered Solids"
edited by A.K. Sen, K.K. Bardhan, B.K. Chakrabarti, (Lecture Note in Physics Vol. 762, Springer-Verlag), (2008) arXiv:0803.0422v1.

\bibitem{Okareview2}
T.~Oka, ``Strong field physics in condensed matter", 
in proceedings of International Conference on Physics in Intense Fields (PIF 2010), arXiv:1102.2482v1.















\bibitem[{\citenamefont{{F.~H.~L.~Essler, H.~Frahm, F.~G\"ohmann, A.~Kl\"umper
  and V.~E.~Korepin}}(2005)}]{Hubbardbook}
\bibinfo{author}{\bibnamefont{{F.~H.~L.~Essler, H.~Frahm, F.~G\"ohmann,
  A.~Kl\"umper and V.~E.~Korepin}}}, \emph{\bibinfo{title}{The One-Dimensional
  Hubbard Model}} (\bibinfo{publisher}{Cambridge}, \bibinfo{year}{2005}).


\bibitem[{\citenamefont{Stafford and Millis}(1993)}]{Stafford1993}
\bibinfo{author}{\bibfnamefont{C.~A.} \bibnamefont{Stafford}} \bibnamefont{and}
  \bibinfo{author}{\bibfnamefont{A.~J.} \bibnamefont{Millis}},
  \bibinfo{journal}{Phys. Rev. B} \textbf{\bibinfo{volume}{48}},
  \bibinfo{pages}{1409} (\bibinfo{year}{1993}).


\bibitem[{\citenamefont{Vasilev and Vitanov}(2004)}]{Vasilev2004}
\bibinfo{author}{\bibfnamefont{G. S.}~\bibnamefont{Vasilev}} \bibnamefont{and}
  \bibinfo{author}{\bibfnamefont{N. V.}~\bibnamefont{Vitanov}},
  \bibinfo{journal}{Phys. Rev. A} \textbf{\bibinfo{volume}{70}},
  \bibinfo{pages}{053407} (\bibinfo{year}{2004}).

\bibitem[{\citenamefont{Fukui and Kawakami}(1998)}]{Fukui}
\bibinfo{author}{\bibfnamefont{T.}~\bibnamefont{Fukui}} \bibnamefont{and}
  \bibinfo{author}{\bibfnamefont{N. }~\bibnamefont{Kawakami}},
  \bibinfo{journal}{Phys. Rev. B} \textbf{\bibinfo{volume}{58}},
  \bibinfo{pages}{16051} (\bibinfo{year}{1998}).

\bibitem[{\citenamefont{Nakamura and Hatano}(2006)}]{NakamuraHatano06}
\bibinfo{author}{\bibfnamefont{Y.}~\bibnamefont{Nakamura}} \bibnamefont{and}
  \bibinfo{author}{\bibfnamefont{N.}~\bibnamefont{Hatano}},
  \bibinfo{journal}{J. Phys. Soc. Jpn.} \textbf{\bibinfo{volume}{75}},
  \bibinfo{pages}{104001} (\bibinfo{year}{2006}).



\bibitem{JeckelmannPRL00}
E.~Jeckelmann, F.~Gebhard, and F.~H.~L.~Essler,
Phys. Rev. Lett. {\bf 85} 3910 (2000).






\bibitem{footnoteE}
Numerically, we find that this holds
for large $U$ as well.
For $U \sim 20$ the difference is at most 5\%.

\bibitem[{\citenamefont{Schollw\"ock}(2005)}]{SchollwoeckReview}
\bibinfo{author}{\bibfnamefont{U.}~\bibnamefont{Schollw\"ock}},
  \bibinfo{journal}{Rev. Mod. Phys.} \textbf{\bibinfo{volume}{77}},
  \bibinfo{pages}{259} (\bibinfo{year}{2005}).

\bibitem{footnoteKE}
$f(\gamma)=(\mbox{K}(v)-\mbox{E}(v))/(\gamma v),\;v=\gamma/\sqrt{1+\gamma^2}$
where K and E are the complete elliptic integral.

\bibitem[{\citenamefont{Tomita and Nasu}(2001)}]{Tomita2001}
\bibinfo{author}{\bibfnamefont{N.}~\bibnamefont{Tomita}} \bibnamefont{and}
  \bibinfo{author}{\bibfnamefont{K.}~\bibnamefont{Nasu}},
  \bibinfo{journal}{Phys. Rev. B} \textbf{\bibinfo{volume}{63}},
  \bibinfo{pages}{085107} (\bibinfo{year}{2001}).




\bibitem{Altshuleroverheating09}
B.~L.~Altshuler, V.~E.~Kravtsov, I.~V.~Lerner, and I.~L. Aleiner,
Phys. Rev. Lett. {\bf 102}, 176803 (2009).  


\bibitem{Hwang2012APS}
K.~Fan, H.~Hwang, M.~Miu, A.~Strikwerda, J.~Zhang, A.~Sternbach, 
X.~Zhang, K.~Nelson, and R.~Averitt, APS March meeting (2012)
http://meetings.aps.org/link/BAPS.2012.MAR.Q17.11.

\bibitem{Philippprivate}
P. Werner, {\it private communication}.


\bibitem{Lenarcic12}
Z.~Lenar\v{c}i\v{c}, and P.~Prelov\v{s}ek, Phys.~Rev.~Lett. {\bf 108}, 196401 (2012).

\bibitem{Uchino12}
S.~Uchino, and N.~Kawakami, Phys.~Rev.~A {\bf 85} 013610 (2012).

\end{thebibliography}

\end{document}